\documentclass[a4paper,11pt]{article}

\usepackage[a4paper,
  left=2.5cm, right=2.5cm,
  top= 3cm, bottom=4cm]{geometry}

\usepackage{amsmath}
\numberwithin{equation}{section}
\usepackage{mathtools}
\usepackage{oldgerm}
\usepackage{amssymb}
\usepackage{bbm}
\usepackage{graphicx}
\usepackage{subcaption}
\usepackage{tikz-cd} 
\usetikzlibrary{matrix,arrows,decorations.pathmorphing}
\usepgfmodule{shapes}

\usetikzlibrary{arrows,decorations.markings}
\usepackage{color}
\usepackage{pgf}
\usepackage{pgffor}

\usepackage{epic}
\usepackage{hyperref}
\usepackage{bbold}

\newcommand{\be}{\begin{equation}}  
\newcommand{\ee}{\end{equation}}  
\newcommand{\bp}{\begin{pmatrix*}[r]}  
\newcommand{\ep}{\end{pmatrix*}}  
\newcommand{\bpp}{\begin{pmatrix}}  
\newcommand{\epp}{\end{pmatrix}}  
\newcommand{\bcd}{\begin{equation}
\begin{tikzcd}}
\newcommand{\ecd}{\end{tikzcd} \end{equation}}

\def\1{\mathbb{1}}

\def\cN{\mathcal{N}}
\def\blue{\textcolor{blue}}
\def\red{\textcolor{red}}

\newcommand{\tr}{\text{Tr}\,}

\begin{document}
\begin{titlepage}
 
\vskip -0.5cm
\rightline{\small{\tt  t16/014}} 
 
\begin{flushright}

\end{flushright}
 
\vskip 1cm
\begin{center}
 
{\huge \bf \boldmath T-branes through 3d mirror symmetry} 
 
 \vskip 2cm
 
Andr\'es Collinucci$^1$, Simone Giacomelli$^1$, Raffaele Savelli$^{2}$, Roberto Valandro$^{3, 4,5}$

 \vskip 0.4cm
 
 {\it  $^1$Physique Th\'eorique et Math\'ematique and International Solvay Institutes,\\ Universit\'e Libre de Bruxelles, C.P. 231, 1050
Bruxelles, Belgium \\[2mm]
 
 $^2$Institut de Physique Th\'eorique, CEA Saclay,\\ Orme de Merisiers, F-91191 Gif-sur-Yvette, France \\[2mm]

$^3$Dipartimento di Fisica, Universit\`a di Trieste,\\ Strada Costiera 11, 34151 Trieste, Italy\\[2mm]
$^4$INFN, Sezione di Trieste, Via Valerio 2, 34127 Trieste, Italy \\[2mm]
$^5$ICTP, Strada Costiera 11, 34151 Trieste, Italy
 }
 \vskip 2cm
 
\abstract{
T-branes are exotic bound states of D-branes, characterized by mutually non-commuting vacuum expectation values for the worldvolume scalars. The M/F-theory geometry lifting D6/D7-brane configurations is blind to the T-brane data. In this paper, we make this data manifest, by probing the geometry with an M2-brane. We find that the effect of a T-brane is to deform the membrane worldvolume superpotential with monopole operators, which partially break the three-dimensional flavor symmetry, and reduce supersymmetry from $\cN=4$ to $\cN=2$. Our main tool is 3d mirror symmetry. Through this language, a very concrete framework is developed for understanding T-branes in M-theory. This leads us to uncover a new class of $\cN=2$ quiver gauge theories, whose Higgs branches mimic those of membranes at ADE singularities, but whose Coulomb branches differ from their $\cN=4$ counterparts.
} 

\end{center}

\end{titlepage}

\tableofcontents

\section{Introduction}
Simple singularities of complex surfaces and semi-simple Lie algebras are both classified by ADE Dynkin diagrams. This coincidence, originally known to mathematicians as the McKay correspondence, has an extremely colorful incarnation in string theory, which not only reproduces it, but gives it a clear meaning.
If one compactifies M-theory or IIA string theory on a K3 surface with a canonical ADE-type singularity, the effective field theory will contain a gauge multiplet for the corresponding Lie algebra. The Cartan components of this multiplet originate from the KK zero modes of the supergravity three-form $C_3$. The roots arise in a more interesting way from the fact that the singularity has spheres of vanishing area that are interconnected in the form of a Dynkin diagram. M2 or D2-branes wrapping such zero size spheres will give rise to massless particles in the effective theory that are charged under the Cartan $U(1)$'s thanks to the minimal coupling $\int_{M2/D2} C_{3}$.

The {\bf A} series of singularities admits another interpretation. The geometry in this case has a circle fibration along which one can reduce M-theory to IIA. The A$_{N-1}$ case gives rise to a system of $N$ coincident D6-branes, which are known to carry an $SU(N)$ gauge group.

This correspondence between singularities and Lie algebras can also be studied from the point of view of a probe M2 or D2-brane that is point-like on the singular K3-surface, and extends over three non-compact directions. In this case, the three-dimensional (3d) field theory exhibits a flavor symmetry corresponding to the singularity in question. This symmetry is not visible in a classical Lagrangian. It can be deduced by exploiting the \emph{3d mirror symmetry} discovered in \cite{Intriligator:1996ex} and further understood in the context of string theory in \cite{Hanany:1996ie,deBoer:1996mp,Porrati:1996xi,deBoer:1996ck,deBoer:1997ka,deBoer:1997kr,Aharony:1997bx}. It can also be deduced directly by introducing the notion of \emph{monopole operators}, and studying their properties as was done in \cite{Kapustin:1999ha,Borokhov:2002ib,Borokhov:2002cg}.

All of these incarnations of the ADE classification have been known for some twenty years. Part of the IIA open string moduli space can be understood in this geometric language. For instance, the IIA system with $N$ coincident D6-branes carries three adjoint-valued Higgs fields $\phi_{\rm D6}^{1,2,3}$. Switching on vevs $\langle \phi_{\rm D6}^i\rangle\neq 0$ will break $SU(N)$ to some subgroup. Usually, such vevs are interpreted as the act of separating the coincident branes, naturally making some of the gluons massive. The M-theory counterpart to this is deforming or resolving the A$_{N-1}$ singularity to a milder singularity.

However, there is a class of vevs that does not admit such a geometric interpretation, vevs such that $[\langle \phi_{\rm D6}^i \rangle, \langle \phi_{\rm D6}^j \rangle] \neq 0$ for some $i, j$. If we complexify two out of the three scalars, then this corresponds to switching on nilpotent vevs for the complexified Higgs, i.e. $\langle \Phi_{\rm D6} \rangle \neq 0$, with $\langle \Phi_{\rm D6} \rangle^p = 0$, for some $p \in \mathbb{Z}$. In this case, the D6-branes are still coincident, but carry only a subgroup of the original $SU(N)$. In the M-theory uplift, the singularity is exactly the same, yet some physical effect is reducing the gauge group. Such vevs were first considered in \cite{Gomez:2000zm} and \cite{Donagi:2003hh}. They were later studied more systematically in \cite{Heckman:2010qv,Cecotti:2010bp} in the context of 7-branes, where they were dubbed `T-branes'. The `T' stands for the fact that the Higgs has an upper triangular vev. The effect is to bind coincident branes together so that they behave as one, and the gauge group is reduced. However, there is no clear proposal to date for their M-theory counterparts. The problem has been analyzed in the related context of F-theory in \cite{Anderson:2013rka,Collinucci:2014taa}, but both these studies need further developments.

Switching on an off-diagonal vev of $\Phi_{\rm D6}$ corresponds in string theory to turning on a coherent state of strings connecting different branes of the stack. These very strings uplift to M2-branes wrapping vanishing cycles of the singular geometry. Therefore, one is led to believe that the uplift of a T-brane is a coherent state of vanishing M2-branes. However, in the absence of a formulation for microscopic M2-branes, we will turn to the 3d perspective of a probe M2-brane that witnesses this effect. This approach will prove very powerful.

From the 3d perspective, a D2-probe in the presence of a stack of D6-branes sees $\langle \Phi_{\rm D6} \rangle$ as a mass for the D2/D6 matter fields, $\tilde Q \langle \Phi_{\rm D6} \rangle Q$. Mass deformations have been studied in the literature, however, only in the case where the mass matrix is diagonalizable. The case of a nilpotent vev (i.e. a T-brane), is very different, and corresponds to a non-diagonalizable mass matrix. This possibility has been pointed out in \cite{Benini:2009qs} for the case of two intersecting D6-branes. It is our goal to study such deformations and their mirror descriptions in depth.

\vskip 3mm
In this paper, we initiate the study of T-branes by probing them with D2-branes. By using mirror symmetry, we learn what a T-brane looks like, when uplifted to M-theory. Switching on a T-brane vev on a stack of $N$ D6-branes corresponds to an off-diagonal mass term on a probe D2-brane, in analogy to the 4d analysis of \cite{Heckman:2010qv}. The mirror of this is a D2 probing and A$_{N-1}$ singularity, with a superpotential deformed by monopole operators. By studying this case we develop a technique that can be extrapolated to D2-branes at any ADE singularity, including the exceptional ones which have no Lagrangian mirror. The main tool we develop for this is what we will refer to as `local 3d mirror symmetry'. It consists in taking a quiver gauge theory, focusing on a single node, ungauging all other nodes, and performing mirror symmetry. This allows us to study the effect of a monopole operator that deforms a single node in terms of an easier mirror theory, finding the low energy effective description, performing a mirror transformation back to the original theory, and finally recoupling the node to the rest of the quiver.

The goal of this paper is to understand what a T-brane looks like in M-theory. By using mirror symmetry, we see how T-brane data gets translated into information on a singular geometry, which is then one simple uplift away from M-theory.

Conversely, this paper introduces a new class of 3d $\cN=2$ theories of a very special kind. These theories have each a natural $\cN=4$ `parent' quiver gauge theory with, as a Higgs branch, a complex surface with an ADE singularity whose Dynkin diagram corresponds to the quiver shape. The $\cN=2$ theory is described by a quiver shape with less nodes than the parent, yet the Higgs branch remains intact. From this, one deduces that the singularity has obstructed blow-up modes, a phenomenon already observed in \cite{Anderson:2013rka}.

The paper is organized as follows: In section \ref{TheA_Nseries}, we review the 3d mirror symmetry for the simplest class of theories, those with $SU(N)$ flavor symmetry. We start with its $\cN=2$ version, and build it up to $\cN=4$. We also explain the string theory realization of the correspondence as a `9-11' flip in M-theory. In section \ref{sec:tbranes}, we review the concept of `T-branes' adapted to D6-branes, and present the issue of understanding their M-theory uplift. We also provide a microscopic interpretation of monopole operators as membranes wrapping vanishing cycles.
In section \ref{sec:simplemirror}, we study T-branes for the {\bf A} series, through mirror symmetry, in the most straightforward way, and find that the effective theories are described by a reduced quiver. In section \ref{sec:o6plane}, we introduce an O$6^-$-plane to the stack of D6-branes that we are probing: We  summarize the mirror dual, which has D$_N$ flavor symmetry, and we discuss the effect of T-branes on the Coulomb branch of the quiver theory. In section \ref{sec:localmirror}, we introduce the technique of `local mirror symmetry': We first test it for the {\bf A} series, for which we already know the result, and then we apply it to the case of minimal T-branes in the {\bf D} and {\bf E} series. In section \ref{sec:conclusions}, we present a summary and an outlook. Finally, in appendix \ref{AppendixA}, we provide several details of the $\cN=4$ mirror map for the D$_4$ theory.

\section{Abelian mirror symmetry: The \texorpdfstring{{\bf A}}{} series}\label{TheA_Nseries}
\subsection{\texorpdfstring{$\cN=2$}{} theory} \label{sec:n2mirror}

Three-dimensional mirror symmetry without Chern Simons terms (the case of interest in this paper) is reviewed in \cite{Aharony:1997bx}. We will briefly explain it here.

The original mirror symmetry is a strong/strong coupling correspondence between two $d=3$, $\mathcal{N}=4$ theories. However, it also exists for $d=3$, $ \mathcal{N}=2$ theories. For the purposes of this article, it will be more useful to proceed anachronistically, by starting from $d=3$, $ \mathcal{N}=2$, and building up to $d=3$, $\mathcal{N}=4$ when necessary. Since $d=3$, $\mathcal{N}=2$ is the dimensional reduction of $d=4$, $\mathcal{N}=1$, we will use the familiar language of the latter.

\paragraph{Theory A:}
The prototype Abelian mirror symmetry has on the one side, what we will call `theory A', an $\cN=2$ theory with the following field content:
\begin{itemize}
\item A $U(1)$ vector multiplet with, as its lowest components, one real scalar $\sigma$ and one photon $A_\mu$. In three dimensions, one can Hodge dualize the photon to a scalar:
\be
d A = \star d \gamma\,.
\ee
The supersymmetrization of this operation corresponds to converting the vector multiplet into a (twisted) chiral multiplet by pairing $\gamma$ with $\sigma$. It is useful to define the exponential of this new complex scalar
\be
V_\pm \sim e^{\pm(i \gamma+\sigma)}\,.
\ee
$V_-$ and $V_+$ are called \emph{monopole operators}. Inserting a monopole operator $V_+$ in the path integral
\be
\int \mathcal{D}[\varphi] \ldots V_+(x) \ldots e^{-S}
\ee
is equivalent to cutting out a small sphere around the spacetime event $x$ and imposing boundary conditions on $A_\mu$ equivalent to having a magnetic monopole of unit charge. Alternatively, in radial quantization on $S^2 \times \mathbb{R}$, acting with $V_\pm$ on the vacuum creates a solitonic state corresponding to a line bundle $\mathcal{O}(\pm 1)$ over the sphere.

\item $N$ pairs of electrons and positrons $(Q_i, \tilde Q^i)$, with $i=1, \ldots, N$, each in a chiral multiplet.

\end{itemize}
The superpotential is zero, $W=0$. This theory can be represented by the following simple quiver:

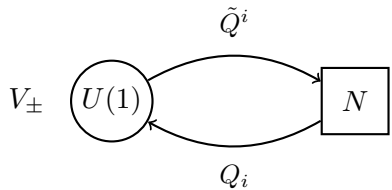
\begin{center}
\begin{tikzpicture}[->,thick, scale=0.8]
  \node[circle, draw, inner sep= 2pt](L1) at (10,0){$U(1)$};
  \node[draw, rectangle, minimum width=25pt, minimum height=25pt](L2) at (14,0){$N$};
\node at(8.6,0) (monopole) {$V_\pm$};
 \path[every node/.style={font=\sffamily\small,
  		fill=white,inner sep=1pt}]
(L1) edge [bend left] node[above=2mm] {$\tilde Q^i$} (L2)
(L2) edge [bend left] node[below=2mm] {$Q_i$}(L1)
;
\end{tikzpicture}
\end{center}
There is a global $U(N) \times U(N)$ symmetry acting on the $Q$ and $\tilde Q$ separately in the $({\bf \bar N, 1})$ and $({\bf 1, N})$ respectively.

The moduli space of vacua splits into two mutually exclusive branches loosely referred to as `Coulomb' and `Higgs' branch. We will refer to these as CB$_A$ and HB$_A$, respectively. They only intersect at their respective origins. The Higgs branch has $\langle \sigma \rangle=0$, and is parametrized by the meson matrix
\be \label{hba1}
{\rm HB}_A: \quad {M_i}^j = Q_i \tilde Q^j\,.
\ee
The only constraint on this matrix is the rank-one condition rk$(M)=1$, i.e.
\be \label{hba2}
{M_i}^{j} {M_k}^{\ell} = {M_i}^{\ell} {M_k}^{j} \qquad \forall \, i, j, k, \ell \:.
\ee

The Coulomb branch consists in vacua with $\langle Q \rangle=\langle \tilde Q \rangle=0$, and $(\gamma, \sigma)$ taking on vevs. The most appropriate coordinates for this branch are the monopole operators $V_\pm$. Na\"ively, it seems redundant to keep both coordinates, since classically $V_+ V_- = 1$. However, there is a one-loop correction, yielding the quantum relation
\be \label{cba}
{\rm CB}_A: \quad V_+ V_- = 0\,.
\ee
The correction comes from the fact that, at the origin of the Coulomb branch, the chiral matter fields become massless, and the na\"ive Wilsonian effective action develops a singularity. It can be derived via heuristic arguments, via a one-loop calculation of the metric of the moduli space, via mirror symmetry, or via a monopole counting argument. 

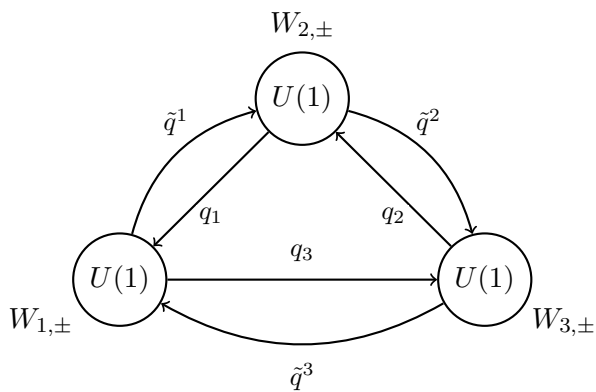
\begin{figure}[h!]
\centering
\begin{tikzpicture}[->, every node/.style={circle,draw},thick, scale=0.8]
  \node(L1) at (-3,-3){$U(1)$};
  \node(L2) at (0,0){$U(1)$};
  \node(L3) at (3,-3){$U(1)$};
\node[draw=none] at (0,1.2) {$W_{2,\pm}$};
\node[draw=none] at (-4.3,-3.7) {$W_{1,\pm}$};
\node[draw=none] at (4.3,-3.7) {$W_{3,\pm}$};

 \path[every node/.style={font=\sffamily\small,
  		fill=white,inner sep=1pt}]
(L1) edge [bend left] node[above=2mm] {$\tilde q^1$} (L2)
(L2) edge  node[below=2mm] {$q_1$}(L1)
(L2) edge [bend left] node[above=2mm] {$\tilde q^2$} (L3)
(L3) edge node[below=2mm] {$q_2$}(L2)
(L3) edge [bend left] node[below=2mm] {$\tilde q^3$} (L1)
(L1) edge node[above=2mm] {$q_3$}(L3)
;

\end{tikzpicture}
\caption{Example of the mirror of $\cN=2$ SQED with $N=3$ flavors}\label{FigN3SQED}
\end{figure}

\paragraph{Theory B:} Now let us define `theory B', which is mirror to theory A. It is described by an Abelian quiver gauge theory, whereby the quiver is shaped like an \emph{affine} Dynkin diagram (see Figure \ref{FigN3SQED}).
\noindent The field content is the following:
\begin{itemize}
\item A $U(1)^N$ gauge group, of which the diagonal subgroup decouples from the rest of the theory. Each node comes with a vector multiplet, of which the lowest components are rewritten as pairs of monopole operators $W_{i, \pm} \sim \exp{\pm (i \gamma_i+\sigma_i)}$.

\item $N$ pairs of fundamental and antifundamental chirals $(q_i, \tilde q^i)$, connecting the nodes.
\item $N$ neutral chiral multiplets $S_i$.
\end{itemize}

\noindent The $\cN=2$ theory comes equipped with the superpotential
\be
W = \sum_{i=1}^N S_i q_i \tilde q^i\,.
\ee

This theory also has a Coulomb and a Higgs branch (CB$_B$ and HB$_B$), which are mutually exclusive. The Higgs branch is parametrized by the following gauge invariant coordinates:
\be \text{$N$ mesons} \quad z_i = q_i \tilde q^i\,, \quad \text{a baryon} \quad B = \prod_{i=1}^N q_i\,, \quad \text{an anti-baryon} \quad \tilde B = \prod_{i=1}^N \tilde q^i\,.
\ee
The F-terms for the $S_i$ set all mesons to zero $z_i = 0$. Hence, we find that the Higgs branch is given by
\be \label{hbb}
{\rm HB}_B: \quad B \tilde B = 0\,,
\ee
where this follows from the definition of the variables. An analysis of the Coulomb branch reveals the following quantum relations:
\be \label{cbb}
{\rm CB}_B: \quad W_{i, +} W_{i, -}  = S_i S_{i-1}\,.
\ee
By inspection, we see that HB$_B$ bears a striking resemblance to CB$_A$ in \eqref{cba}, and CB$_B$ to HB$_A$ in \eqref{hba1} and \eqref{hba2}. This prompts the following identifications:
\begin{eqnarray}
V_+ &\longleftrightarrow& B \qquad V_- \longleftrightarrow \tilde B\\ \nonumber
{M_i}^i &\longleftrightarrow& S_i \qquad {M_i}^{i-1} \longleftrightarrow W_{i, -} \qquad {M_{i-1}}^i \longleftrightarrow W_{i, +}
\end{eqnarray}
This correspondence is essentially the content of mirror symmetry. The branches get exchanged, and quantum corrected relations (for the Coulomb branches) get rewritten in terms of quantum exact classical F-terms (for the Higgs branches). The Higgs branch is protected from quantum corrections in $\cN=4$ theories, but also in Abelian $\cN=2$ theories.

\subsection{\texorpdfstring{$\cN=4$}{} theory}

Having setup the $\cN=2$ mirror symmetry, it is now easy to obtain a version with enhanced $\cN=4$ supersymmetry. We essentially keep the same theories `A' and `B', but making some mild modifications.

\paragraph{Theory A:}
\begin{itemize}
\item The $Q$ and $\tilde Q$ are now paired up as hypermultiplets. 
\item The $U(1)$ vector multiplet described by the monopole operators $V_\pm$ is completed to an $\cN=4$ vector multiplet by pairing it up with an $\cN=2$ chiral multiplet of lowest component~$\Phi$. 
\item Finally, the superpotential is fixed by $\cN=4$ supersymmetry to be
\be
W = \sum_{i=1}^N \tilde Q^i \Phi Q_i\,.
\ee
\end{itemize}
This superpotential constrains the meson matrix to be traceless, Tr$M=0$. The new quiver is the following:

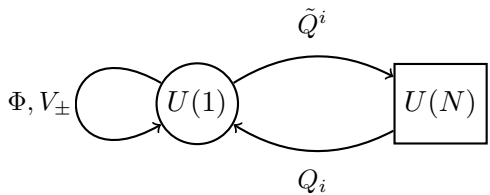
\begin{center}
\begin{tikzpicture}[->,thick, scale=0.8]
  \node[circle, draw, inner sep= 2pt](L1) at (10,0){$U(1)$};
  \node[draw, rectangle, minimum width=30pt, minimum height=30pt](L2) at (14,0){$U(N)$};
 \path[every node/.style={font=\sffamily\small,
  		fill=white,inner sep=1pt}]
(L1) edge [bend left] node[above=2mm] {$\tilde Q^i$} (L2)
(L2) edge [bend left] node[below=2mm] {$Q_i$}(L1)
(L1) edge [loop, in=210, out=150, looseness=8] node[left] {$\Phi, V_\pm$}(L1)
;
\end{tikzpicture}
\end{center}
The equation for the Coulomb branch CB$_A$ is modified due to the fact that the flavors can acquire mass whenever $\Phi$ has a vev. It turns out that the quantum exact equation is
\be
{\rm CB}_A: \quad V_+ V_- = \Phi^N\,.
\ee
This is the equation of the A$_{N-1}$ singularity.

\paragraph{Theory B:} 

\begin{figure}[ht!] 
\centering
\begin{tikzpicture}[->, every node/.style={circle,draw},thick, scale=0.8]
  \node(L1) at (-3,-3){$U(1)$};
  \node(L2) at (0,0){$U(1)$};
  \node(L3) at (3,-3){$U(1)$};

 \path[every node/.style={font=\sffamily\small,
  		fill=white,inner sep=1pt}]
(L1) edge [bend left] node[above=2mm] {$\tilde q^1$} (L2)
(L2) edge  node[below=2mm] {$q_1$}(L1)
(L2) edge [bend left] node[above=2mm] {$\tilde q^2$} (L3)
(L3) edge node[below=2mm] {$q_2$}(L2)
(L3) edge [bend left] node[below=2mm] {$\tilde q^3$} (L1)
(L1) edge node[above=2mm] {$q_3$}(L3)

(L2) edge [loop, out=120, in=60, looseness=4] node[above] {$\varphi_2, W_{2,\pm}$} (L2)
(L1) edge [loop, out=240, in=170, looseness=4] node[left=1mm] {$\varphi_1, W_{1,\pm}$} (L1)
(L3) edge [loop, out=295, in=0, looseness=4] node[right=1.5mm] {$\varphi_3, W_{3,\pm}$} (L3)
;

\end{tikzpicture}
\caption{Example of the mirror of $\cN=4$ SQED with $N=3$ flavors}\label{fig:n4quiver}
\end{figure}
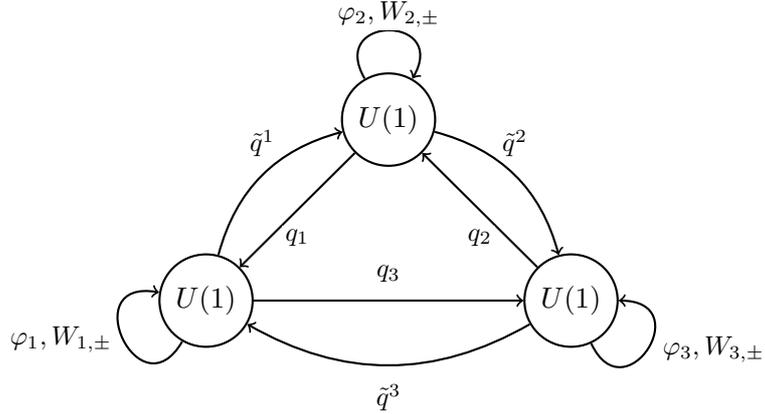

Here, we only make one change. A chiral field $\Psi$ is added, and the superpotential is augmented to the following:
\be
W=\sum_{i=1}^N S_i q_i \tilde q^i- \Psi \sum_{i=1}^N S_i\,.
\ee
$\Psi$ is massive, and its F-term imposes the constraint $\sum_{i=1}^N S_i=0$. Note that this matches the tracelessness constraint for the meson matrix on the A-side, and alters the geometry of the Coulomb branch CB$_B$.
After integrating out $\Psi$, we can solve for its F-term by rewriting the $S_i$ as differences of chiral multiplets
$S_i = \varphi_i-\varphi_{i+1}$, giving rise to the quiver diagram in figure \ref{fig:n4quiver}.

\noindent The $(q_i, \tilde q^i)$ pairs now form hypers, and the $\varphi_i$ are naturally combined with the vectors of each node into $\cN=4$ vector multiplets. Note, that if we keep $\Psi$ in the Lagrangian, the F-terms for the $S_i$ will impose
\be
q_i \tilde q^i = \Psi\,.
\ee
The new Higgs branch equation will then be
\be
{\rm HB}_B: \quad B \tilde B = \Psi^N
\ee
i.e. the A$_{N-1}$ singularity. It is therefore natural to postulate the correspondence:
\be
\Phi \leftrightarrow - \Psi\,.
\ee

\subsection{Brane picture}
Let us now briefly review the M-theory embedding of the $\cN=4$ theories, and the mirror correspondence. The latter can be understood as a chain of dualities from IIA to itself, namely a TST chain. But it is easiest to understand it as a `9-11' flip, i.e. starting form an M-theory configuration, and choosing two different available circles to reduce to IIA. The following diagram summarizes the idea:

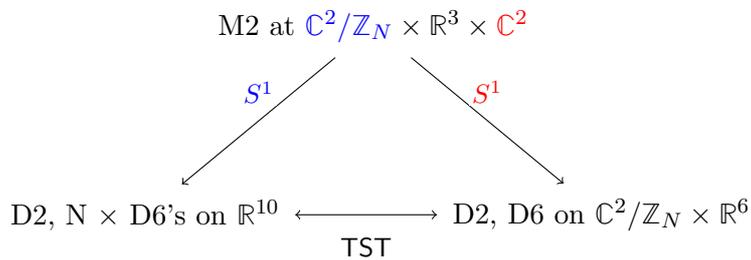
\begin{center}
\begin{tikzpicture}[node distance = 2cm, auto, inner sep=2mm]
\node  (M) at (0,0) {M2 at $ \blue{\mathbb{C}^2/\mathbb{Z}_N} \times \mathbb{R}^3 \times \red{\mathbb{C}^2}$};
\node  (IIB) at (-3,-2.5)  {D2, N $\times$ D6's on $\mathbb{R}^{10}$};
\node  (IIA) at (3,-2.5) {D2, D6 on $\mathbb{C}^2/\mathbb{Z}_N \times \mathbb{R}^6$};
 \path[every node/.style={font=\sffamily\small,
  		fill=white,inner sep=1pt}]

(M) edge[->] node[midway, above=5.5pt]{$\red{S^1}$} (IIA)
(M) edge[->] node[midway, above=5.5pt]{$\blue{S^1}$} (IIB)
(IIA) edge[<->] node[sloped, inner sep=3mm, midway, below]{TST} (IIB);
(IIB) edge  (IIA);

\end{tikzpicture}
\end{center}
\noindent In M-theory, we have an M2-brane probing an A$_{N-1}$ singularity, and filling out the $\mathbb{R}^3$. So both the orbifold in blue and the $\mathbb{C}^2$ in red are transverse to it. 
\paragraph{Theory A:}
The orbifold has a natural circle fibration in it, it can be understood as a limiting geometry in a family of $N$-centered Taub-NUT spaces. Reducing along this, the blue circle, we get a D2 that probes $N$ D6-branes in flat spacetime. The setup is summarized by the following:

\begin{center}
$
\begin{array}{c|cccccccccc}
 & 0 & 1 & 2 & 3 & 4 & 5 & 6 & 7 & 8 & 9 \\
N\times \text{D6}  & \times & \times & \times & \times & \times & \times & \times &  &  &   \\
\text{D2} & \times & \times & \times &  &  &  &  &  &  &  \\
  \multicolumn{4}{c}{} \hspace*{\dimexpr2\tabcolsep} & 
    \multicolumn{4}{@{}c@{}}{\underbrace{\hspace*{\dimexpr7\tabcolsep+6\arrayrulewidth}\hphantom{0123}}_{\text{decoupled hyper}}} & 
        \multicolumn{3}{@{}c@{}}{\underbrace{\hspace*{\dimexpr4\tabcolsep+6\arrayrulewidth}\hphantom{012}}_{\begin{subarray}{l} \text{vector m.} \\  \\ \Phi,\, \sigma,\, A_\mu \end{subarray} }} 
\end{array}
$
\end{center}

The theory has $\mathcal{N}=4$ supersymmetry, but we will use the $\mathcal{N}=2$ language to describe the multiplets. The field content is the following:
\begin{itemize}
\item A decoupled hypermultiplet (two chiral mutiplets) containing the scalars $\phi^{I=3, \ldots, 6}$ that represent movement along the D6-brane.
\item An $\mathcal{N}=4$ vector multiplet that breaks into an $\mathcal{N}=2$ chiral and a vector multiplet as follows
\be
(\Phi \equiv \phi^7+i \phi^8); \quad (\sigma \equiv \phi^9, A_\mu)\,.
\ee
As explained before, we can construct monopole operators $V_\pm \sim \exp{\pm(\sigma+i\gamma)}$, where $\gamma$ is the dual photon.

\item A hyper that breaks into two oppositely charged multiplets $(Q^i, \tilde Q_i)$, with $i=1, \ldots, N$, coming from D2/D6 stretched strings.

\end{itemize}

\paragraph{Theory B:} 
We can choose a different M-theory circle along which to reduce to IIA, by making a simple observation. The $\mathbb{C}^2$ in red, in our main diagram, can be written as a single-centered Taub-NUT geometry. This is simply a circle fibration over $\mathbb{R}^3$, where the fiber collapses over one point. Reducing along that circle gives IIA with a single D6-brane. In this case, we are left with a D2-brane probing $\mathbb{C}^2/\mathbb{Z}_N$, in the presence of a single D6-brane. The latter will not give us any interesting information in our analysis, so we will drop it from now on. The theory of a D2 probing an orbifold singularity is a well-understood one, and it gives rise exactly to the quiver gauge theory we referred to as the `theory B'. We can now explain the theory as follows:
\begin{itemize}
\item The D2 breaks up into fractional branes, each represented by a node of the quiver. Each fractional brane is actually a D4-brane wrapping a vanishing sphere of the singular geometry. It comes with its vector multiplet, here broken up into a chiral $\varphi_i$, and a vector $W_{i, \pm}$. The $\varphi_i$ can be thought of as the positions of the fractional branes along two non-compact directions.
\item Two adjacent fractional branes have open strings stretched between them, giving rise to the $(q_i, \tilde q^i)$ fields.
\item The superpotential $W = \sum (\varphi_i-\varphi_{i+1}) q_i \tilde q^i$ expresses the fact that, if two fractional branes move apart along the remaining non-compact directions, the stretched strings acquire mass.
\end{itemize}
  
\section{T-branes and their M-theory counterparts} \label{sec:tbranes}
\subsection{IIA perspective}

Here, we briefly review the concept of the so-called `T-branes', adapted to our case of interest in IIA string theory. 

A stack of $N$ D6-branes will naturally host a $U(N)$ gauge group, and its field content will carry three adjoint Higgs fields $\phi^I_{\rm D6}$, with $I=7, 8, 9$, corresponding to the three transverse directions. Whenever anyone of the latter acquires a vev, the gauge group will break to the subgroup of $U(N)$ that commutes with the $\langle \phi^I_{\rm D6} \rangle$. Typical vevs for the $\phi^I_{\rm D6}$ are diagonalizable, and the eigenvalues are interpreted as the positions of the constituent D6-branes. Naturally, as branes are separated, the stretched strings that accounted for the non-Abelian gauge group become massive, thereby explaining the breaking.
It could happen, however, that the vevs for the three Higgses are not simultaneously diagonalizable. In that case, one can no longer interpret the Higgsing as separating the branes. For the purposes of this paper, we single out one of the three transverse directions, say $\phi_{\rm D6}^9$, and pair up the other two into a complex field $\Phi_{\rm D6} \equiv \phi_{\rm D6}^7+i \phi_{\rm D6}^8$.

We will define a T-brane as a stack of D6-branes where $\Phi_{\rm D6}$ has a nilpotent vev, i.e. $\langle \Phi_{\rm D6} \rangle^p = 0$ for some $p$. This implies that all the eigenvalues of $\Phi_{\rm D6}$ are zero, and the branes are still very much coincident. Nevertheless, the gauge group is broken to a subgroup. For example, on a stack of $4$ D6-branes,
\be \label{minimalorbit} \langle \Phi_{\rm D6} \rangle = \begin{pmatrix} 0 & 0 & 0 & 0\\ 1 & 0 & 0 & 0 \\0 & 0 & 0 & 0 \\ 0 & 0 & 0 & 0 \end{pmatrix} \ee
the unbroken gauge group is $U(1) \times U(2)$.
Physically, two of the four branes are forming a bound state with a unique center of mass, and the other two are forming a $U(2)$ sub-stack. These bound states were first studied in \cite{Gomez:2000zm} and \cite{Donagi:2003hh}. Later, in \cite{Heckman:2010qv,Cecotti:2010bp} the scope of the analysis was broadly expanded to cases of non-perturbative 7-branes.

The example given here \eqref{minimalorbit} is what is known as a \emph{minimal nilpotent orbit}. It corresponds to the gauge orbit of this matrix under adjoint $U(4)$ transformations. One could also have matrices with two, and three ones in the superdiagonal. These correspond to higher nilpotent orbits. In this paper, we will mostly focus on minimal orbits.

The fact that these non-trivial vevs have no geometric interpretation in terms of brane positions has a counterpart in the M-theory uplift. As explained in the previous section, IIA in the presence of several D6-branes uplifts to a purely geometric background known as the multi-centered Taub-NUT space. Essentially, the M-theory circle is non-trivially fibered over the transverse $\mathbb{R}^3$, and it collapses above the locus of each D6-brane. When the D6-branes coincide, these `centers' where the fiber collapses coalesce, forming an orbifold $\mathbb{C}^2/\mathbb{Z}_N$ singularity.\footnote{Strictly speaking, one gets an ALF space. By taking the limit where the asymptotic radius of the M-theory circle goes to infinity, one gets the $\mathbb{C}^2/\mathbb{Z}_N$ orbifold.}

\subsection{M-theory perspective}

Since switching on diagonalizable vevs for $\Phi_{\rm D6}$ corresponds to moving the D6-branes apart, in M-theory, this data translates into a deformation of the singularity. However, for T-branes, the singularity remains intact, even though we expect the gauge group to break. How is this breaking seen in M-theory?

This is a question that has barely been addressed, and to our knowledge there are only two proposals for studying this phenomenon in the related context of F-theory \cite{Anderson:2013rka,Collinucci:2014taa}. For the time being, both proposals consist in sophisticated mathematical constructions that might appropriately encode T-branes into the singular geometry in M/F-theory. However, their physical meanings need further development.

In principle, the uplift of a T-brane to M-theory can be characterized as follows: In general, switching on a vev for a worldvolume field $\Phi_{\rm D6}$ on a D6-brane corresponds to turning on a coherent state of strings in the spectrum corresponding to $\Phi_{\rm D6}$. Strings that go from one brane to itself will uplift to metric moduli in M-theory. However, strings stretched between different branes on the stack uplift to M2-branes wrapping an $S^2$ that is a circle fibration over the interval connecting the two branes. When the branes coincide, the $S^2$ shrinks to zero size, and the membrane gives rise to an effective massless particle.
Therefore switching on a vev for an off-diagonal Higgs corresponds precisely to a coherent state of M2-branes that wrap the sphere corresponding to the root of the Lie algebra along which the vev points. This heuristic picture, as convincing as it may be, requires a mathematical formalism in order to actually compute things.

In this paper, we will approach T-branes by probing them with D2-branes. We will see that we will gain a clear view on these phenomena, and most of all, computational power. We will start with the case of coincident D6-branes, which uplift to $\mathbb{C}^2/\mathbb{Z}_N$ singularities. But we will learn enough from that simple class of examples to be able to study the rest of the ADE series.

What we will show is that, on the mirror side, a D2-brane probing the mirror of a T-brane has a monopole operator deforming its Lagrangian. Schematically, this is summarized as follows:
\be
\langle \Phi_{\rm D6} \rangle=\begin{pmatrix} 0&0\\m&0 \end{pmatrix} \qquad \Rightarrow \quad \Delta W_{\rm D2}= m Q_1 \tilde Q^2 \quad \xleftrightarrow{\rm \quad mirror \quad} \quad \Delta W_{\rm D2}= m W_{2,+}\,.
\ee

We claim that these deformations by monopole operators gives an M-theory definition of what a T-brane is without reference to IIA string theory. The point is that, even though the A-theory description of a T-brane as an off-diagonal mass is simple, and its infrared theory does describe the M2-brane, it is only available in this form for the {\bf A} and {\bf D} series. On the other hand, the mirror description of a T-brane as a superpotential deformation by a monopole operator, although less straightforward, is more universal, and can be used to describe the {\bf E} series.

The core of this paper will therefore consist in studying quiver gauge theories deformed by monopole operators.

\subsection{String theory interpretation of monopole operators}
So far we have defined monopole operators in field theory. It is however useful to gain some intuition about them by finding their string theoretic interpretation. In this section, we find such an interpretation for magnetic monopoles on fractional D2-branes at singularities. We will describe it in two ways.

\begin{paragraph}{Operator-state correspondence}
In this paragraph, we will use the operator-state correspondence to show that monopole operators map to states of D2-branes wrapping vanishing spheres. 

One way to define a monopole operator $W_+(x)$ is as a disorder operator that enforces a singularity on the 3d gauge field at the space-time point $x$, such that, for any two-sphere surrounding it, we have
\be
\int_{S^2} F = 2 \pi \cdot 1\,.
\ee

Since the theory of the M2-brane is the IR fixed point of the D2 theory, we can apply the operator state correspondence, and map 
\be
\mathbb{R}^3 \mapsto S^2 \times \mathbb{R}\,.
\ee
From the perspective of radial quantization on $\mathbb{R}^3$, time is the radial direction, and the two-spheres of equal radius correspond to spacelike slices. Placing a monopole operator at the origin gets mapped to preparing a particle state at time $\tau = -\infty$, with magnetic charge $\int_{S^2} F = 2 \pi$. Let us now think about our fractional D2, which is a D4-brane wrapping a vanishing $\mathbb{P}^1$. Its Wess-Zumino worldvolume coupling to the $C_3$ form now becomes a source for induced D2 charge:
\be
S_{\rm WZ} = \mu_{\rm D4} \int_{S^2 \times \mathbb{R} \times \mathbb{P}^1} F \wedge C_3 = 2\pi \mu_{\rm D4} \int_{\mathbb{R}\times \mathbb{P}^1} C_3\,.
\ee
Hence, inserting a monopole operator at the origin of $\mathbb{R}^3$ corresponds to creating a magnetic D-particle at $\tau = -\infty$ from a D2 wrapping an exceptional cycle.

\end{paragraph}

\begin{paragraph}{Open membranes}
In the previous paragraph, we appealed to the operator state correspondence in order to see a D-particle. In this paragraph, we will see this even more directly. 

Let us consider a D4-brane, with an open D2-brane ending on it. The fact that this configuration is possible has been established in \cite{Townsend:1996em,Argurio:1997nh,Argurio:1997gt,Argurio:1997qv,Strominger:1995ac}. The argument is as follows: The full IIA supergravity action plus worldvolume theories of the host D4 and open D2-branes contains the Chern-Simons and Wess-Zumino terms
\begin{eqnarray} 
S &=& \tfrac{1}{2} \int_{X^{10}} \left( \tilde G_4 \wedge \star \tilde G_4 + B_2 \wedge G_4 \wedge G_4\right) + \mu_4 \int_{D4} (F_2+\imath_4^* B_2) \wedge \imath_4^* C_3 + \mu_2 \int_{D2}\imath_2^* C_3\,, \nonumber \\
{\rm with} \quad \tilde G_4 &=& d C_3-C_1 \wedge H_3\,, 
\end{eqnarray}
where $F$ is the DBI field-strength on the D4-brane, and $\imath_p^*$ represents the pullback onto the worldvolume of a $p$-brane. In this setup, we will impose $H_3 = 0$, so we can freely use the `unimproved' field strength $G_4 = dC_3$. The presence of the D4-brane implies a sourced Bianchi identity
\be 
d G_4 = \mu_4 \delta_5\,,
\ee
where, in general, by $\delta_k$ we mean the $k$-form that is Poincar\'e dual to a $(10-k)$-dimensional object. We can write the equations of motion for $C_3$
as follows:
\begin{eqnarray}
d \left( \star G_4 +G_4 \wedge B_2 \right)&=& \mu_4 \delta_5 \wedge (F_2+B_2)+ \mu_2 \delta_7 \nonumber \\
\Rightarrow d \star G_4 &=&  \mu_4 \delta_5 \wedge F_2+ \mu_2 \delta_7 
\,.
\end{eqnarray}
Now we can integrate both sides of the equation on an $S^7$ that intersects the D2-brane at one point, and the D4-brane at the $S^2$ that surrounds the boundary of the D2-brane:
\be
0=  \mu_4 \int_{S^2 } F_2+ \mu_2\,.
\ee
This implies that $F_2$ must take on the profile of a codimension three defect on the D4-brane:
\be
d F_2 = -\frac{\mu_2}{\mu_4} \delta^{(4)}_3\,,
\ee
where $\delta^{(4)}_3$ is a threeform defined on the D4 that is Poincar\'e dual to the boundary of the open D2.

Now that we have analyzed the case in flat ten dimensions, the analysis can be repeated in the case where both the D4 and the D2 are wrapping an exceptional $\mathbb{P}^1$. The answer remains the same: \emph{The insertion of a D2 ending on the D4 induces a 3d instanton on the 3d worldvolume of the D4}. In other words, we will have $d F_2 = \frac{\mu_2}{\mu_4} \delta_3$ in the 3d theory.
Therefore, the insertion of a D2-brane that ends on the point $x$ in $\mathbb{R}^3$ has the same effect as inserting $W_+(x)$ in the path integral. Hence, \emph{we identify the open D2 with a monopole operator}.

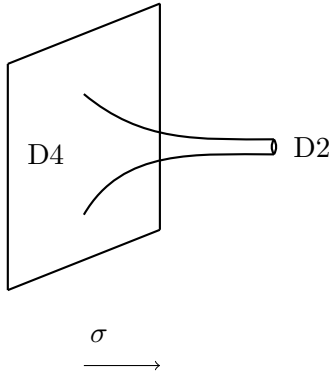
\begin{figure}[ht!]
\begin{center}
\begin{tikzpicture}
\draw[thick] (0,0) -- (2,0.8);
\draw[thick] (0,-3) -- (2,-2.2);
\draw[thick] (0,0) -- (0,-3);
\draw[thick] (2,0.8) -- (2,-2.2);
\draw[thick] (1,-.4) to [out=320, in=180] (3.5, -1);
\draw[thick] (1,-2) to [out=60, in=180] (3.5, -1.2);
\draw[thick] (3.5, -1) to [bend left] (3.5, -1.2);
\draw[thick] (3.5, -1) to [bend right] (3.5, -1.2);
\draw[->] (1,-4)--(2,-4);
\node at (1.2, -3.6) {$\sigma$};
\node at (.5, -1.2) {D4};
\node at (4, -1.1) {D2};

\end{tikzpicture}
\end{center}
\caption{When $g_s$ is turned on, a D2 ending on a D4 becomes a smooth funnel shape.}
\label{fig:funnel}
\end{figure}
\end{paragraph}

How is this related to the state-operator correspondence explained before? Or more directly, can we see that this is equivalent to creating a D2-particle state? The answer is a resounding `yes'. It can be shown that the supersymmetric solution for the 3d instanton solution requires the real scalar $\sigma$ to acquire a profile of the form 
$$\sigma \sim \frac{q}{r} \qquad {\rm with} \quad q = \frac{\mu_2}{\mu_4}\,.$$ 
Since $\sigma$ represents a transverse coordinate to the fractional D2-brane, (i.e. wrapped D4-brane), this means that the open D2 is pulling on the D4, stretching it into a funnel shape, as depicted in figure \ref{fig:funnel}. The induced worldvolume metric of the D4-brane is now:
\be
ds^2 = \left(1+\frac{1}{r^4} \right) dr^2+r^2 d\Omega_{2}^2\,,
\ee
which is conformally equivalent to both $\mathbb{R}^3$ and $S^2 \times \mathbb{R}$. The point is that now we can alternate between the two pictures that characterize a monopole operator, simply by changing the choice of the direction we call `Euclidean time': 
\begin{itemize}
\item If we choose a Cartesian coordinate, say the vertical axis, then the funnel looks like a disturbance localized in space and time, from the perspective of the D4. In other words, it looks like an instanton created by the monopole operator.
\item If we choose the direction $r$ to be our Euclidean time, then the system looks like a D2-brane wrapping a vanishing $\mathbb{P}^1$ that appears as a magnetic particle in 3d, whereby the spacelike slices of spacetime grow with time.
\end{itemize}

This geometry allows us to see the operator-state correspondence fully embedded in string theory. The point of view that a monopole operator creates a D2-particle state bolsters our claim that off-diagonal strings stretched between D6-branes should uplift in M-theory to M2-branes wrapping vanishing cycles, since such strings appear on the D2 as off-diagonal mass terms that are mirror to monopole operators.

\section{T-branes and mirror symmetry: The \texorpdfstring{\bf A}{} series} \label{sec:simplemirror}
\subsection{T-branes as deformations by monopole operators}

From the perspective of the worldvolume theory on a D2 probing D6-branes, the Higgs field on a stack of D6-branes appears as a background field, or a coupling in three dimensions. Starting with just $N$ D6-branes, if we switch on a vev $\langle \Phi_{\rm D6}\rangle = {\rm diag}(0, 0, \ldots, 0, m, -m)$, this will correspond to moving the last two branes apart symmetrically, leaving the D2 brane in the middle. We therefore expect the two flavors to gain equal and opposite masses
\be
W = \Phi_{\rm} \sum_{i=1}^N (Q_i \tilde Q^i) + m (Q_{N-1} \tilde Q^{N-1}-Q_{N} \tilde Q^{N})\,.
\ee
In the infrared, we are left with $N-2$ flavors, and hence the new quantum corrected equation for the Coulomb branch will be
\be
V_+ V_- = \Phi^{N-2}\,.
\ee
This perfectly matches the fact that the M-theory singularity has been deformed to a milder one. On the B-side, these mass terms are sent to the following terms
\be
m (Q_{N-1} \tilde Q^{N-1}-Q_{N} \tilde Q^{N}) \longrightarrow m (S_{N-1}-S_{N})\:.
\ee
Now, the F-terms for the Higgs branch are modified as follows:
\be
q_i \tilde q^i = \Psi\,\quad i \neq N-1, N\,, \quad q_{N-1} \tilde q^{N-1} = \Psi-m\,, \quad q_{N} \tilde q^{N} =\Psi+m\,,
\ee
from which we find
\be
B \tilde B = \Psi^{N-2} (\Psi-m) (\Psi+m)\,.
\ee
Hence, the singularity has been deformed. 

This takes care of diagonalizable masses. The main subject of this paper is to study what happens when we turn on non-diagonalizable masses. For example, take
\be\label{acccc}
W = \Phi_{\rm} \sum_{i=1}^{N} (Q_i \tilde Q^i) + m Q_{N-1} \tilde Q^N\,.
\ee
Clearly, two chiral flavors (that do not fit into the same hyper) become massive. One might suspect that the Coulomb branch equation would account for that by lowering the power of $\Phi$ by two. However, the effective theory after integrating out the massive flavors is qualitatively different from the class of theories we have been considering:
\be \label{effectiveA}
W_{\rm eff} = \Phi_{\rm} \sum_{i=1}^{N-2} (Q_i \tilde Q^i) - \frac{\Phi^2}{m} P \tilde P\,,
\ee
where $P \equiv Q_{N}, \tilde P \equiv \tilde Q^{N-1}$. This off-diagonal mass term breaks the $\cN=4$ to $\cN=2$. Now there are less flavors, but one of them has a coupling quadratic in $\Phi$. We expect that the Coulomb branch equation remains qualitatively unmodified as follows:
\be
V_+ V_- = \Phi^{N-2} \times \left(-\frac{\Phi^{2}}{m}\right)\,.
\ee
In order to confirm this, one needs to repeat the calculations of \cite{Borokhov:2002cg} in this new context.

Let us now investigate what happens on the mirror side. The off-diagonal mass operator we have introduced gets mapped to a monopole operator
\be
m\, Q_{N-1} \tilde Q^N \longrightarrow m W_{N, +}\,.
\ee
Here, it becomes very difficult to say what happens as a result of this deformation. $W_{N, +}$ is not a fundamental field in 
the UV, where the theory is weakly coupled, so we cannot simply differentiate the superpotential with respect to it. There are 
several strategies around this problem. One of them is to make the mirror map of the effective theory \eqref{effectiveA}. We do 
this as follows:

First, we start with the $\cN=2$ mirror symmetry, as explained in section \ref{sec:n2mirror}, but for a theory with $N-1$ flavors.
\be
W_A = 0 \longrightarrow W_B = \sum_{i=1}^{N-1} S_i (q_i \tilde q^i)\,.
\ee
Now we supplement the A-side with its superpotential \eqref{effectiveA}, and map each term to the B-side, 
\be\label{nilpot}
W_B = \sum_{i=1}^{N-1} S_i (q_i \tilde q^i) - \Psi \sum_{i=1}^{N-2} S_i - \frac{\Psi^2}{m} S_{N-1}\,.
\ee
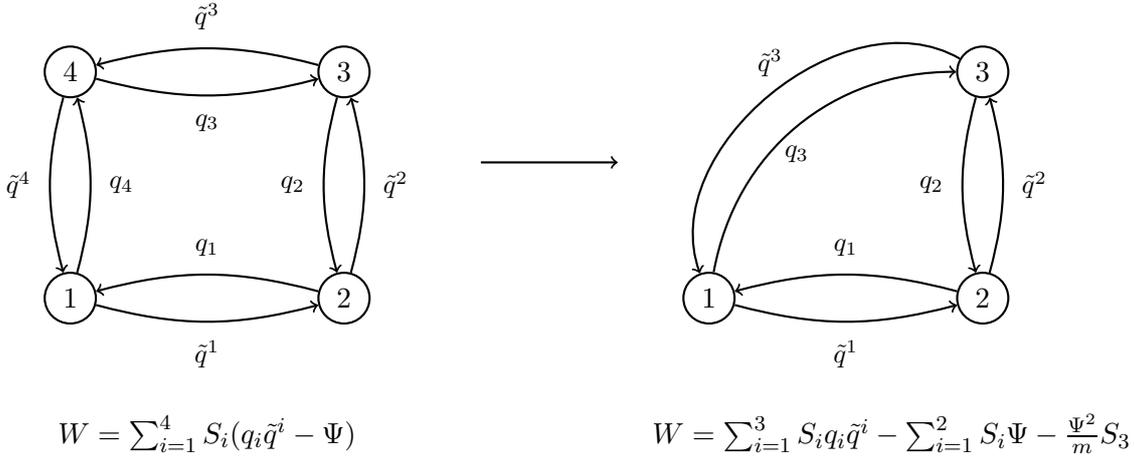
\begin{figure}
\centering
\begin{tikzpicture}[->, every node/.style={circle,draw},thick, scale=0.6]

  \node(M1) at (-17,-3){$1$};

 \node(M2) at (-11,-3){$2$};
    \node(M3) at (-11,2){$3$};
        \node(M4) at (-17,2){$4$};
            \node[draw=none](T1) at (-14,-6) {$W = \sum_{i=1}^4 S_i (q_i \tilde q^i-\Psi)$};

  \node(L1) at (-3,-3){$1$};
  \node(L2) at (3,-3){$2$};
    \node(L3) at (3,2){$3$};
    \node[draw=none](T2) at (1,-6) {$W = \sum_{i=1}^3 S_i q_i \tilde q^i-\sum_{i=1}^2 S_i \Psi -\frac{\Psi^2}{m} S_3$};

\draw[=>] (-8,0)--  (-5,0);

 \path[every node/.style={font=\sffamily\small,
  		fill=white,inner sep=1pt}]
(M1) edge [bend right=15] node[below=2mm] {$\tilde q^1$} (M2)
(M2) edge [bend right=15]  node[above=2mm] {$q_1$}(M1)
(M2) edge [bend right=15] node[right=2mm] {$\tilde q^2$} (M3)
(M3) edge [bend right=15] node[left=2mm] {$q_2$}(M2)
(M3) edge [bend right=15] node[above=2mm] {$\tilde q^3$} (M4)
(M4) edge [bend right=15] node[below=2mm] {$q_3$}(M3)
(M4) edge [bend right=15] node[left=2mm] {$\tilde q^4$} (M1)
(M1) edge [bend right=15] node[right=2mm] {$q_4$}(M4)

(L1) edge [bend right=15] node[below=2mm] {$\tilde q^1$} (L2)
(L2) edge [bend right=15]  node[above=2mm] {$q_1$}(L1)
(L2) edge [bend right=15] node[right=2mm] {$\tilde q^2$} (L3)
(L3) edge [bend right=15] node[left=2mm] {$q_2$}(L2)
(L3) edge [bend right=70] node[above=2mm] {$\tilde q^3$} (L1)
(L1) edge [bend left=40] node[below=2mm] {$q_3$}(L3)
;

\end{tikzpicture}
\vskip -2.8cm
\caption{Example of the A$_{3}$-theory with a minimal T-brane. The effect of the T-brane on the quiver is to remove the corresponding node and substitute the arrow ending and starting from that node with new arrows that connect the adiacent nodes (by abuse of notation the new quarks are also denoted as $q_3,\tilde{q}^3$).}
 \label{fig:N=4->N=3 first}
\end{figure}
This is an $\cN=2$ theory described by a quiver diagram with $N-1$ nodes instead of $N$ with bifundamentals and $N-1$ neutral chiral multiplets which we again call $q_i,\tilde{q}_i$ and $S_i$ respectively (see figure \ref{fig:N=4->N=3 first} for an example). We will recover this result again in section \ref{sec:localmirrorabelian} via a more general method. Let us analyze its Higgs branch to see what kind of singularity we get. The F-terms for the $S_i$ give the following equations:
\be
q_i \tilde q^i = \Psi\,, \quad i \neq N-1 \,, \quad q_{N-1} \tilde q^{N-1} = \frac{\Psi^2}{m}\,.
\ee
From this, we create again the following invariants:
\be
B \equiv \prod_{i=1}^{N-1} q_i\,,  \quad \tilde B \equiv \prod_{i=1}^{N-1} \tilde q^i\,, 
\ee
for which we deduce the relation
\be
B \tilde B = \Psi^{N-2} \left(\frac{\Psi^2}{m} \right) \sim \Psi^N\,.
\ee
As expected from the M-theory picture, the A$_{N-1}$ singularity stays undeformed! 

In order to test this correspondence in a non-trivial way, we will match HB$_A$ with CB$_B$. On the A-side, the effective superpotential \eqref{effectiveA} obtained by integrating out the massive fields, gives the following F-term equation for $\Phi$
\be
\sum_{i=1}^{N-2} Q_i \tilde Q^i - 2\,\frac{\Phi}{m} P \tilde P=0\,.
\ee
The F-term equations for the various electrons and positrons tell us that the HB and CB are still disjoint outside the origin. Hence, on HB$_A$, the meson matrix gets a partial tracelessness condition
\be
\sum_{i=1}^{N-2} {M_i}^i = 0\,.
\ee
Otherwise, the full meson matrix still satisfies the rank one condition just as before, the main difference being that it is smaller by one row and one column. 

Let us now see what the B-side shows. We now have the $S_i$ satisfying a partial sum condition on the Coulomb branch ($\langle \Psi \rangle=0$)
\be
\sum_{i=1}^{N-2} S_i = 0\,.
\ee
At each node, there is a pair of monopoles $W_{i, \pm}$ like before, except that there is one node less. In order to find the 
equations governing the CB geometry, we repeat an argument in \cite{Aharony:1997bx}: At each node, there is a topological $U(1)$ 
symmetry sending $W_\pm \mapsto e^{\pm i\alpha} W_\pm$, which means that the CB must be a circle fibration over a space. However, 
since the Higgs branch of that Abelian theory is invariant under this $U(1)$, it must be the case that the CB and HB intersect 
at a fixed point of the $U(1)$ group action. This means that the circle fiber collapses to a point. This implies a geometry of the following form:
\be
W_{i,+} W_{i,-} =S_i S_{i-1}\,.
\ee
To understand this, note, that whenever either $S_i$ or $S_{i-1}$ are zero, a part of the Higgs branch becomes unobstructed. 
This equation tells us that we have a $\mathbb{C}^*$-fibration over the $(S_i, S_{i-1})$-plane that collapses over the origin.
In conclusion, we see that HB$_A$ matches CB$_B$. 

The case of a general nilpotent mass term can be treated along the same lines. Let us consider a mass matrix in Jordan form (with 
nonzero elements under the diagonal). A Jordan block of size k corresponds to adding to the superpotential the following terms: 
\be\label{jordangen} m\sum_{i=1}^{k-1}Q_i\tilde{Q}^{i+1}.\ee 
The massless fields now are $Q_{k}$, $\tilde{Q}^1$ and $Q_j$, $\tilde{Q}^j$ with $j>k$. Below the scale $m$ we then get a $U(1)$ 
theory with $N-k+1$ flavors. Using the F-term equations 
\be mQ_i+\Phi Q_{i+1}=0;\quad m\tilde{Q}_i+\tilde{Q}_{i-1}\Phi=0,\ee 
we can see that when we integrate out massive fields we generate the superpotential term 
\be(-1)^{k-1}\frac{\Phi^k}{m^{k-1}}Q_k\tilde{Q}^1,\ee 
which is mapped in the mirror theory (a quiver with $N-k+1$ nodes) to $-\Psi^kS_{N-k+1}/m^{k-1}$. Repeating the analysis performed before 
for the case $k=2$, we get to the conclusion that the singularity is still A$_{N-1}$. Notice that adding the mass terms 
(\ref{jordangen}) corresponds to turning on superpotential terms involving monopole operators at $k-1$ 
consecutive nodes in the mirror side.

\subsection{Is the singularity frozen?}

The fact that the effective quiver for the mirror of a minimal T-brane looks like an A$_{N-2}$ Dynkin diagram, but its moduli 
space describes an A$_{N-1}$ singularity has a very interesting consequence. It implies that one vanishing sphere is obstructed 
from being blown up. More precisely, blowing up a sphere would correspond to adding a real FI term to its corresponding node. The 
loss of a node, however, implies the loss of a $U(1)$ factor, which in turn means we have one less real FI parameter at our 
disposal. This dovetails nicely with the observation of \cite{Anderson:2013rka} that T-branes obstruct blow-ups of singularities.

We would now like to make a comment about the complex FI terms on the B-side (superpotential terms linear in the fields $S_i$), 
which are related to deformations of the singularity as was explained before. In the original $\mathcal{N}=4$ theory we have $N$ 
gauge groups and hence $N$ complex FI terms. However, supersymmetry requires their sum to vanish so the truly independent parameters 
are $N-1$. One equivalent way to see this is as follows: The superpotential for the $\mathcal{N}=4$ theory can be written in the form 
\be W=\sum_{i=1}^N S_i(q_i \tilde q^i) - \Psi \sum_{i=1}^{N} S_i \:.\ee 
We can now turn on linear superpotential terms for all the $S_i$ fields, $\sum_i a_iS_i$, and with a redefinition of the field 
$\Psi$ we can set the sum of the $a_i$ to zero. 
On the other hand, once we have turned on the nilpotent mass term the superpotential becomes (\ref{nilpot}), and since the 
field $\Psi$ now appears quadratically in the superpotential, we end up generating new superpotential terms by shifting it. 
Consequently, we are no longer allowed to reabsorb a combination of the complex FI parameters $a_i$ with a redefinition of the 
fields. Since in the process we lost one gauge node, we conclude that we still have $N-1$ independent FI parameters, or 
equivalently $N-1$ deformation parameters. 

A related observation is the following fact noticed in \cite{Heckman:2010qv}: Given a nilpotent mass matrix ${\bf m}$, we can obtain a 
diagonalizable one by adding its hermitian conjugate ${\bf m}^{\dagger}$. The sum of the two mass terms does not break extended supersymmetry 
anymore, since ${\bf m}+{\bf m}^{\dagger}$ trivially commutes with its hermitian conjugate. We can imagine turning on the above mass deformation 
in two steps: first we consider the matrix ${\bf m}$ only, which breaks $\mathcal{N}=4$, and then turn on the second mass term in the resulting 
theory. In the IR we expect to recover the $\mathcal{N}=4$ theory associated with ${\bf m}+{\bf m}^{\dagger}$. 

We are now in the position to explicitly check this: Let us consider again (\ref{acccc}). By integrating out the massive field 
and extracting the mirror we get (\ref{nilpot}). The hermitian conjugate mass matrix leads of course to the term 
$m^* Q_{N} \tilde Q^{N-1}$ on the A side. Turning on this term corresponds, on the B side, to adding in (\ref{nilpot}) the FI term $m^* S_{N-1}$. 
The F-term equations can be solved by setting to zero $\Psi$ and all the mesons except $q_{N-1} \tilde q^{N-1}$, whose vev should be 
equal to $-m^*$. This higgses the neighbouring gauge groups to the diagonal combination and by expanding the superpotential around this 
vev we generate a mass term for $S_{N-1}$. Integrating out massive fields we are left with 
\be W=\sum_{i=1}^{N-2} S_i(q_i \tilde q^i) - \Psi \sum_{i=1}^{N-2} S_i,\ee
which corresponds precisely to the $\mathcal{N}=4$ quiver of type $A_{N-3}$. One can also see the deformation of the A$_{N-1}$ 
singularity explicitly. In fact the F-terms resulting from adding $m^*S_{N-1}$ to (\ref{nilpot}) are $q_i\tilde q^i = \Psi$ 
for $i=1,\dots,N-2$ and $mq_{N-1}\tilde q^{N-1}=\Psi^2-|m|^2$. After defining the baryons as usual, $B=q_1\dots q_{N-1}$ and 
$\tilde B=\tilde q^1 \dots\tilde q^{N-1}$, one obtains the relation
\begin{equation}
  B \, \tilde B \sim \Psi^{N-2}(\Psi^2-|m|^2).
\end{equation}

The case of Jordan blocks of arbitrary size can be treated analogously. The only difference is that the superpotential 
terms associated with the mass matrix ${\bf m}^{\dagger}$ now involve massive fields. 
The case of a Jordan block of size three will be enough to illustrate this point. The superpotential term related to ${\bf m}$ is 
\be mQ_1\tilde{Q}^2+mQ_2\tilde{Q}^3.\ee 
So $Q_1$, $Q_2$, $\tilde{Q}^2$ and $\tilde{Q}^3$ are all massive and we have the following F-term equations: 
\be m Q_1+\Phi Q_2 = m Q_2+\Phi Q_3 = 0; \quad m\tilde{Q}^2+\tilde{Q}^1\Phi = m\tilde{Q}^3 + \tilde{Q}^2\Phi = 0.\ee 
As was explained in the previous section, when we integrate out massive fields we get an $\mathcal{N}=2$ effective theory with 
the superpotential term 
\be\frac{\Phi^3}{m^2}Q_3\tilde{Q}^1.\ee
The superpotential related to ${\bf m}^{\dagger}$ is instead 
\be m^*Q_2\tilde{Q}^1+m^*Q_3\tilde{Q}^2\ee 
and both terms involve massive fields ($Q_2$ and $\tilde{Q}^2$ respectively). Using the above F-term equations we can rewrite 
this as 
\be\label{jordandiag} -2\frac{m^*}{m}\Phi Q_3\tilde{Q}^1.\ee
We conclude that turning on the superpotential term associated with ${\bf m}^{\dagger}$, corresponds to turning on (\ref{jordandiag}) 
in the $\mathcal{N}=2$ effective theory, which in turn is mapped to a term of the form $\Psi S_1$ in the mirror theory. 

We find that in general, the mass term related to ${\bf m}^{\dagger}$ is mapped in the mirror to terms of the form 
$\Psi^{k-2}S_{i}$ (where $k$ is the size of the Jordan block). We could also consider mass terms related to matrices of the 
form $({\bf m}^{\dagger})^n$, which turn out to be mapped in the mirror to the terms $\Psi^{k-n-1}S_i$ (with $n<k$), or also terms 
related to diagonal mass matrices, which instead are mapped to $\Psi^{k-1}S_i$.
Repeating the computation of the previous paragraphs, one immediately sees that all these terms do deform the singularity. 
The outcome is that the superpotential terms $S_i$ and $\Psi^nS_i$ (with $n$ smaller than the size of the corresponding Jordan 
block), which are all related to ``diagonalizable'' completions of the mass matrix, correspond to deformations of the singularity. 
In total we always have $N$ such terms but one of them can be removed with a shift of $\Psi$. We conclude that we always have 
$N-1$ deformation parameters.

\section{T-branes and mirror symmetry: The \texorpdfstring{{\bf D}}{} series} \label{sec:o6plane}
\subsection{Basic setup}
So far, we have only discussed the simplest case of 3d mirror symmetry. In this section, we will introduce another simple class of mirror pairs. They are summarized with the following diagram

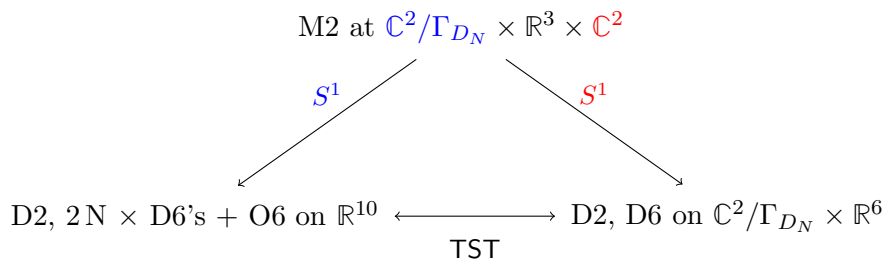
\begin{figure}[ht!]
\begin{center}
\begin{tikzpicture}[node distance = 2cm, auto, inner sep=2mm]
\node  (M) at (0,0) {M2 at $ \blue{\mathbb{C}^2/\Gamma_{D_N}} \times \mathbb{R}^3 \times \red{\mathbb{C}^2}$};
\node  (IIB) at (-3.5,-2.5)  {D2, 2\,N $\times$ D6's $+$ O6 on $\mathbb{R}^{10}$};
\node  (IIA) at (3.5,-2.5) {D2, D6 on $\mathbb{C}^2/\Gamma_{D_N} \times \mathbb{R}^6$};
 \path[every node/.style={font=\sffamily\small,
  		fill=white,inner sep=1pt}]

(M) edge[->] node[midway, above=5.5pt]{$\red{S^1}$} (IIA)
(M) edge[->] node[midway, above=5.5pt]{$\blue{S^1}$} (IIB)
(IIA) edge[<->] node[sloped, inner sep=3mm, midway, below]{TST} (IIB);
(IIB) edge  (IIA);

\end{tikzpicture}
\end{center}
\end{figure}

\noindent Here, $\Gamma_{D_N}$ is the discrete subgroup of $SU(2)$ of order $4 (N-2)$ that leads to a D$_N$ singularity. On the A-side, we have an O6-plane with $N$ D6-branes and $N$ image-D6-branes on top of it. There is a D2/image-D2-pair sitting on top of the O6-plane. The 3d gauge group is $Sp(1)$. The flavor symmetry, corresponding to the gauge group of the D6-branes, is $SO(2N)$.

\begin{paragraph}{Theory A:} The theory A is defined as an $SU(2)$ gauge theory with $N$ flavours $(Q^a_i,\tilde{Q}^j_b)$, with $a=1,2$ the gauge index and $i,j=1,...,N$ the flavor index. The associated quiver diagram is the following:
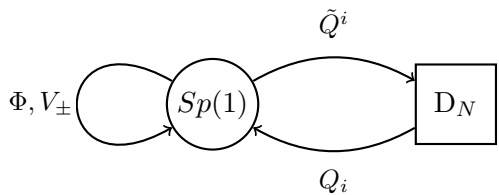
\begin{center}
\begin{tikzpicture}[->,thick, scale=0.8]
  \node[circle, draw, inner sep= 2pt](L1) at (10,0){$Sp(1)$};
  \node[draw, rectangle, minimum width=30pt, minimum height=30pt](L2) at (14,0){D$_N$};
 \path[every node/.style={font=\sffamily\small,
  		fill=white,inner sep=1pt}]
(L1) edge [bend left] node[above=2mm] {$\tilde Q^i$} (L2)
(L2) edge [bend left] node[below=2mm] {$Q_i$}(L1)
(L1) edge [loop, in=210, out=150, looseness=8] node[left] {$\Phi, V_\pm$}(L1)
;
\end{tikzpicture}
\end{center}
The $\cN=4$ theory has the following superpotential
\begin{equation}\label{N=4SuperpotDnA}
W = \sum_{i=1}^{N} Q_i^a \Phi_a^b \tilde{Q}_b^i  \:.
\end{equation}
One can see this theory as the quotient of a $U(2)$ gauge theory with $2N$ flavors (i.e. where now $i,j=1,...,2N$). The O$6$-plane imposes an orientifold projection through an involution that acts as
\begin{equation}
 \tilde{Q}_a^i \mapsto i \gamma_{ab} Q^b_j \Gamma^{ji} \qquad \qquad Q^a_i \mapsto - i \gamma^{ab} \tilde{Q}^j_b \Gamma_{ji} \:,
\end{equation}
where $\gamma=\sigma_2$ (the second Pauli matrix) and $\Gamma$ is the $2N\times 2N$ matrix that in block-form can be written as
\begin{equation}
 \Gamma = \left(\begin{array}{cc}
 0 & {\bf 1}_{N} \\ {\bf 1}_{N} & 0 \\
 \end{array}\right)\:.
\end{equation}
Before the projection, the mesonic matrix is given by
\begin{equation} \label{mesonsu2}
 {M_i}^j \equiv Q_i^a \tilde{Q}_a^j \:.
\end{equation}
After imposing the projection, the quarks with $i,j=N+1,...,2N$ can be written in terms of the ones with $i,j=1,...,N$. The mesonic matrix parametrizing the Higgs branch becomes then constrained. In particular it can be written in block form as
\begin{equation}\label{MorietifoldForm}
 M = \left(\begin{array}{cc}
 A & B \\ C & -A^T \\
 \end{array}\right)\:,
\end{equation}
where $A$ is a generic $N\times N$ matrix, while $B,C$ are antisymmetric $N\times N$. Because of its definition \eqref{mesonsu2}, this $2N\times 2N$ matrix has rank 2. 
The three F-terms for $\Phi$ tell us that, in addition, $M$ must satisfy $M^2=0$. Counting also the three conditions coming from the D-terms, the complex dimension of HB$_A$ is then $4N-6$.
Even if it is not immediate, this form of the meson matrix can be mapped by an isomorphism to an antisymmetric (rank 2) meson matrix.

Let us study the Coulomb branch, i.e. $M=0$ and $\Phi\neq 0$. The D-term condition $[\Phi,\Phi^\dagger]=0$ imposes that $\Phi=\varphi T^3$, where $T^3$ is the Cartan generator of the $Sp(1)$ algebra. A vev for such a field breaks $Sp(1)$ to $U(1)$. Along this branch we can define two monopole operators
$U_\pm$ that are charged under the topological symmetry corresponding to the Cartan $U(1)$ photon. Analogously to the A$_{N-1}$ case, we can write down the following quantum relation (that takes into account the fact that $\varphi$ controls the mass of both charged hypermultiplets and vector multiplets):
\begin{equation}\label{noninvDNsingEq}
  U_+ U_- = \varphi^{2N-4} \:.
\end{equation}
Both $U_\pm$ and $\varphi$ are not gauge invariant, since they transform under the $Sp(1)$ Weyl symmetry: $U_+\leftrightarrow U_-$ and $\varphi\rightarrow-\varphi$. We then define the gauge invariant coordinates on CB$_A$ as $u\equiv\frac{i}{2}\varphi(U_+ - U_-)$, $v\equiv \frac12(U_+ + U_-)$ and $w\equiv \varphi^2$. Plugging these relations into \eqref{noninvDNsingEq}, the equation defining the (complex) two dimensional CB$_A$ becomes
\begin{equation}\label{DNsingCBA}
  u^2+ w v^2 = w^{N-1} \:,
\end{equation}
i.e. the $D_{N}$ singularity.

\end{paragraph}

\begin{center}
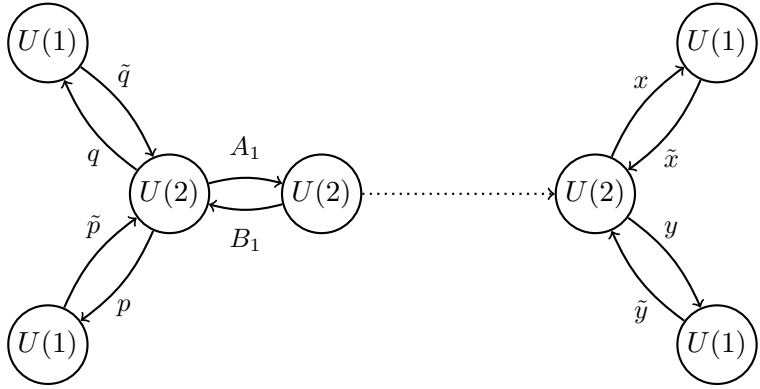
\begin{figure}[ht!] 
\centering
\begin{tikzpicture}[->, every node/.style={circle,draw},thick, scale=0.8]
  \node[inner sep=1.7](L1) at (-10,2.5){$U(1)$};
  \node[inner sep=1.7](L2) at (-10,-2.5){$U(1)$};
  \node[inner sep=1.7](V1) at (-8,0){$U(2)$};
  \node[inner sep=1.7](V2) at (-5.5,0){$U(2)$};
  \node[inner sep=1.7](VN+1) at (-1,0){$U(2)$};
  \node[inner sep=1.7](L3) at (1,-2.5) {$U(1)$};
    \node[inner sep=1.7](L4) at (1,2.5) {$U(1)$};

 \path[every node/.style={font=\sffamily\small,
  		fill=white,inner sep=1pt}]
(L1) edge [bend left=15] node[above=2mm] {$\tilde q$} (V1)
(V1) edge [bend left=15] node[below=2mm] {$q$}(L1)
(L2) edge [bend left=15] node[above=2mm] {$\tilde p$} (V1)
(V1) edge [bend left=15] node[below=2mm] {$p$}(L2)
(V1) edge [bend left=15] node[above=2mm] {$A_1$} (V2)
(V2) edge [bend left=15] node[below=2mm] {$B_1$} (V1)
(V2) edge [dotted] (VN+1)
(L3) edge [bend left=15] node[below=2mm] {$\tilde y$} (VN+1)
(VN+1) edge [bend left=15] node[above=2mm] {$y$}(L3)
(L4) edge [bend left=15] node[below=2mm] {$\tilde x$} (VN+1)
(VN+1) edge [bend left=15] node[above=2mm] {$x$}(L4)

;

\end{tikzpicture}
\caption{D$_N$ quiver.}\label{DNquiver}
\end{figure}
\end{center}

\begin{paragraph}{Theory B:}
Since the theory A has a D$_{N}$ flavor symmetry, it should come as no surprize that, for the theory B, we have a quiver gauge theory with the quiver shaped like a D$_{N}$ affine Dynkin diagram, with non-Abelian nodes in the middle line (see Figure \ref{DNquiver}).

The arrows of the quiver represent bifundamental chirals, as the diagram shows.\footnote{In our convention, the arrows that go from a non-Abelian node to an Abelian one represent column vectors.} In order not to clutter the figure, we did not include the adjoint chiral multiplets, so we list them here. 
\begin{itemize}
\item The four external, Abelian nodes have each a neutral chiral multiplet. Starting from the upper left in clockwise orientation, these are $\phi_q, \phi_x, \phi_y, \phi_p$. In $\cN=2$ language, each of these chirals is accompanied by a vector multiplet.
\item Similarly, each non-Abelian node in the middle horizontal line has an adjoint chiral superfield $\Psi_1\,, \ldots \Psi_{N-3}$, accompanied by a $U(2)$ gauge multiplet.
\end{itemize}
The $\cN=4$ theory has the following superpotential
\begin{eqnarray}
W &=& {\rm Tr}\big(\left(\Psi_1-\mathbb{1} \phi_q \right) q \tilde q+\left(\Psi_{N-3}-\mathbb{1} \phi_x \right) x \tilde x+\left(\Psi_{N-3}-\mathbb{1} \phi_y \right) y \tilde y +\left(\Psi_1-\mathbb{1} \phi_p \right) p \tilde p\big) \nonumber\\
&+& \sum_{i=1}^{N-4} \left(B_i \Psi_i A_i-A_i \Psi_{i+1} B_i \right) 
\end{eqnarray}

The Higgs branch HB$_B$ is described by gauge invariant combinations of the quark fields subject to relations coming from the F-terms for the fields $\Psi_i$ and $\phi_{p,q,x,y}$ \cite{Lindstrom:1999pz,Borokhov:2003yu}. When $N$ is even, the three invariants
\begin{eqnarray}
 z &\equiv& - \tilde{q}p\tilde{p}q\:, \\
 y &\equiv& 2 \tilde{p}A_1\cdots A_{N-4}x \tilde{x}B_{N-4}\cdots B_1 p + (-z)^{N/2-1}\:, \\
 x &\equiv& 2 \tilde{q}A_1\cdots A_{N-4}x \tilde{x}B_{N-4}\cdots B_1 p \tilde{p}q
\end{eqnarray}
satisfy the equation
\begin{equation}\label{DNsingHBB}
  x^2+ z y^2 = z^{N-1} \:,
\end{equation}
that matches with the equation \eqref{DNsingCBA} defining CB$_A$ under the map $z\leftrightarrow w$, $y \leftrightarrow v$ and $x \leftrightarrow u$.
When $N$ is odd, the invariants  satisfying the equation \eqref{DNsingHBB} are defined in a different way \cite{Lindstrom:1999pz}.

The Coulomb branch CB$_B$ is described by Weyl invariant combinations of the fields $\Psi_i$, $\phi_q, \phi_x, \phi_y, \phi_p$ and monopole operators with definite charges under the topological symmetries relative to each node of the D$_N$ quiver. 
They satisfy quantum relations due to the fact that the quarks acquire mass when the $\Psi_i$, $\phi_q, \phi_x, \phi_y, \phi_p$ get a non-zero vev (see equations \eqref{GaiottovAvB} in Appendix \ref{AppendixA} \cite{Bullimore:2015lsa}).  
The D-term conditions on the $U(2)$ adjoint scalars $\Psi_i$ allow only vev proportional to the diagonal $U(1)$ generator or to the Cartan generator of $SU(2)$. 
Along the Coulomb branch the gauge group is broken to $U(1)^{2N-3}$ (where one $U(1)$ has decoupled) and hence its complex dimension is equal to $4N-6$ as expected from mirror symmetry. 

The mirror map between HB$_A$ and CB$_B$ is quite involved (and will be described in Appendix~\ref{AppendixA} for the D$_4$ case). 
Analogously to the A$_{N-1}$ case, the diagonal mesons (when $M$ has the form \eqref{MorietifoldForm}) are mapped into combination of the scalar fields $\Psi_i$, $\phi_q, \phi_x, \phi_y, \phi_p$,\footnote{By a careful analysis, one can see that the only component of $\Psi_i$ that appears in these combinations is the one relative to the diagonal $U(1)$ of the corresponding $U(2)$ node.} while the off-diagonal mesons are mapped to monopole operators with R-charge equal to one.\footnote{Since we are in a $\cN=4$ 3d theory, these operators sit in the same multiplet of conserved currents correspoding to the roots of the non-Abelian flavor symmetry.}
Both the off-diagonal mesons and the monopole operators have definite charges with respect to the Cartan generator of the $SO(2N)$ flavor group. For each set of topological charges we have one monopole operator (with R-charge equal to one), that will be sent to the meson matrix element with the same charges.
One can check that the rank 2 condition on the meson matrix $M$ translates to the quantum relations involving the monopole operators and the scalar fields that define CB$_B$ (see Appendix \ref{AppendixA}). 
\end{paragraph}

\subsection{T-branes}

Having introduced the mirror symmetry for the D$_N$ case, we can now set out to study the effect of T-branes. The problem is substantially complicated by the fact that both the A and B theory are non-Abelian. In this section, we will examine the deformation induced on the F-terms of theory A  by a general T-brane. In this way, we can infer the consequences on the HB$_A$, and thus, by mirror symmetry, we can deduce how CB$_B$ looks like after the deformation. In section \ref{sec:localmirrorde}, instead, we will study the effective field theories on the B-side, and concentrate on HB$_B$ for the simplest class of T-branes corresponding to minimal nilpotent orbits.

On the A-side, a T-brane is described by a deformation of the superpotential \eqref{N=4SuperpotDnA} by a term of the form $\Delta W = {\rm Tr}(m M)$, where $M$ is the meson matrix defined in \ref{mesonsu2}, and $m$ is a \emph{nilpotent} $2N\times 2N$ mass matrix. For the present analysis, it is more convenient to choose a basis where both $m$ and $M$ are antisymmetric. This is always possible, since both matrices are in the adjoint of $SO(2 N)$. 

Since the flavor symmetry is partially broken by the T-brane, it is natural to expect that part of HB$_A$ is lifted, namely, the part that does not commute with $m$. In other words, one might na\"ively expect conditions of the form $[m,M]=0$, which can be translated via mirror symmetry to a statement about non-conserved currents in the adjoint of the quantum flavor group. However, the conditions $[m,M]=0$ are not sufficient to satisfy all of the supersymmetric constraints for (the vev of) $M$. At least in theories with flavor algebras in the {\bf A} and {\bf D} series, supersymmetric vacua are such that $mM=Mm=0$.\footnote{There are some other universal conditions on $M$, which do not depend on the T-brane specified by $m$. They are tr$M=0$ for the {\bf A} series and $M^2=0$ for the {\bf D} series.} Let us see why this is the case.

Firstly, the fact that $m$ commutes with $M$ says that $m|_{{\rm Im}(M)}:{\rm Im}(M)\to {\rm Im}(M)$. Since $m$ is nilpotent by hypothesis, $m|_{{\rm Im}(M)}$ has a non-trivial kernel, and therefore 
\be\label{inequal}
{\rm rk}\,(mM) < {\rm rk}\,(M)\,.
\ee

For theories in the {\bf A} series, the gauge invariant F-terms involving the T-brane deformation read
\begin{eqnarray}\label{IReom}
(\1_{N}t+m)M&=&0\,,\nonumber\\
M(\1_{N}t+m)&=&0\,,
\end{eqnarray}
where $t$ should be thought of as the $\Phi$ ($\Psi$) field of section \ref{TheA_Nseries}, if we use theory A (B) to describe the IR fixed point. Clearly the difference of these two equations gives us $[m,M]=0$. Moreover, in HB$_A$/CB$_B$, which are the branches we are interested in here, $M$ is non-vanishing and ${\rm rk}(M)=1$. The latter condition, due to \eqref{inequal}, immediately implies $mM=0$. Then, equations \eqref{IReom} force the vev of $t$ to vanish as well in this branch.

For theories in the {\bf D} series, things are more tricky. In this case, the gauge invariant F-terms involving the T-brane deformation read
\be\label{GaugInvEomDn}
P+mM=0\,,
\ee
where $P$ is a generator of the chiral ring, \emph{independent} of the others, which has the property of being a \emph{symmetric} matrix in the basis where $m$ and $M$ are antisymmetric. While a description of the operator $P$ in theory B is unknown, and would require a formulation of mirror symmetry for non-Abelian theories, we can still present it in theory A. Following \cite{Borokhov:2003yu}, we package the $N$ flavors $\{Q,\tilde{Q}\}$ into the following doublet of $2N$-vectors
\be
\psi^a=\frac{1}{\sqrt 2}\left(\begin{array}{c}Q^a-\epsilon^{ab}\tilde{Q}_b\\i[Q^a+\epsilon^{ab}\tilde{Q}_b]\end{array}\right)\qquad a,b=1,2\,,
\ee
with $\epsilon^{12}=1$. In this basis, one has
\begin{eqnarray}
M_{ij}&=&\psi^a_i\epsilon_{ab}\psi^b_j\,,\nonumber\\
P_{ij}&=&\psi^a_i\epsilon_{ab}\Phi^b_c\psi^c_j\,,
\end{eqnarray}
where $P$ is symmetric in $i,j$ because the $Sp(1)$-adjoint field $\Phi$ satisfies $\epsilon_{ab}\Phi^b_c=\epsilon_{cb}\Phi^b_a$.

Equations \eqref{GaugInvEomDn} imply $[m,M]=0$. However, here ${\rm rk}(M)=2$, which does not necessarily imply $mM=0$. For any non-trivial T-brane $m$, indeed, there are choices of $M$ such that ${\rm rk}(mM)=1$. Nevertheless, it turns out that all of them lead to vacua which break supersymmetry. In the description of theory A, such breaking occurs through violation of the D-terms for $\Phi$:\footnote{Recall that for the {\bf A} series there are no D-terms for $\Phi$.}
\be\label{DtermsPhi}
[\Phi,\Phi^\dagger]=0\,.
\ee
Indeed, using the D-terms for $\psi$, i.e. $\psi^\dagger_I\psi_I=|\psi|^2\1_2$, equations \eqref{GaugInvEomDn} lead to the following condition on the restriction of $m$ to ${\rm Im}(M)={\rm Span}\{\psi^1,\psi^2\}$:
\be
m_{ij}\psi_j^a=\Phi^a_b\psi^b_i\,.
\ee
Therefore, since $\Phi$ is traceless, the nilpotency of $m$ determines a violation of the D-terms \eqref{DtermsPhi}, except when $\Phi=0$, which means $m|_{{\rm Im}(M)}=0$, or equivalently $mM=0$.

We have seen that, upon any T-brane deformation of the theories in both the {\bf A} and {\bf D} series, supersymmetry still rules out branches of the moduli space where both the mesons $M$ and $\Phi$ are non-vanishing, as was the case for the undeformed theories. 

Now we can use the power of mirror symmetry to learn what happens to the Coulomb branch on the dual quiver side. Mirror symmetry maps the diagonal elements of the meson matrix $M$ to appropriate linear combinations of the adjoint scalars along the $U(1)$'s, including the diagonal $U(1)$'s of the various $U(2)$ nodes. On the other hand, the off-diagonal terms of $M$ are mapped to various monopole operators charged under the appropriate topological $U(1)$ symmetries of the various quiver nodes. Roughly, when $M$ has the form \eqref{MorietifoldForm},
\be
M \quad \longrightarrow \quad \mathcal{M} \equiv \begin{pmatrix} \sum_i c^1_i \Phi_i & W_{2, +} & \cdots \\ \ W_{2, -} & \sum_i c^2_i \Phi_i & \cdots \\ \vdots &\vdots&\ddots  \end{pmatrix}\,.
\ee
The F-term conditions for $M$ are now mapped to F-term conditions for this matrix $\mathcal{M}$ defined above:
\be m M= M m=0 \quad \longrightarrow \quad m \mathcal{M}= \mathcal{M} m=0\,.
\ee
Now the classical equations from the A-side have given us highly non-trivial quantum equations on the B-side that relate Lagrangian variables to monopole operators, telling us how  CB$_B$ is partly lifted. For example, if we take $\mathcal{M}$ and $m$ in the form \eqref{MorietifoldForm}, a minimal T-brane with non-zero $m_{1,2}=-m_{N+2,N+1}$ implies that all the monopole operators and scalar fields in the rows $2$ and $N+1$ and columns $1$ and $N+2$ must vanish.

However, the lack of a detailed $\cN=2$ mirror map for the {\bf D} series, prevents us from deducing what goes wrong in theory B with vacua where $mM\neq0$. It would be interesting to fill in this gap, and thus be able to generalize the lesson to theories in the {\bf E} series.

\section{Local mirror symmetry} \label{sec:localmirror}

\subsection{General strategy}
We have learned from the previous sections that the mirror of a T-brane on the A-side (i.e. a D2-brane probing stacks of D6-branes), is the quiver gauge theory of a D2-brane probing an affine ADE singularity, whereby the superpotential is deformed by monopole operators $\Delta W \sim \sum_i m_i W_{i, +}$, where the $m_i$ are the `mass' parameters on the A-side.

In the {\bf A} and {\bf D} series, there is a perturbative theory A to define the T-brane. However, in the {\bf E} series, the analog of the A-side corresponds to the dimensional reduction of the Minahan-Nemenchansky theories, which are non-Lagrangian. Hence, for these cases, we will define a T-brane directly on the B-side, as a deformation of the quiver gauge theory by monopole operators. In order to study the effect of such deformations, however, we need to develop a new strategy, as we cannot simply study the theory A. In this section, we present this new strategy first via the examples of the {\bf A} and {\bf D} series, and finally apply it to the {\bf E} series.

The idea we propose is the following: Given a quiver gauge theory with a deformation by a monopole operator $\Delta W = m_i W_{i, +}$ 
corresponding to the i-th dual photon, we focus on this i-th node by taking the gauge couplings at neighbouring nodes to be very 
small. In this way we can ignore their dynamics and consider the i-th node as a theory with a single gauge group. The bifundamental 
multiplets are now simply interpreted as fundamentals of this gauge group. Then we consider the mirror of the i-th node ``in isolation'', 
which is in general a linear quiver with off-diagonal mass terms. We can then integrate out massive fields, attempt to extract 
the mirror, and finally reinsert this resulting theory into the original quiver. The key fact is that this theory and the original 
one are equivalent in the IR. 

To summarize, this is our procedure for treating a T-brane:
\begin{enumerate}
\item Define a T-brane for a Dynkin quiver gauge theory, called the `theory B', where the $\cN=4$ superpotential is supplemented 
by a term $\Delta W = m_i W_{i, +}$, where $m_i$ is a parameter, and $W_{i, +}$ is the monopole operator charged under the 
topological $U(1)$ of the i-th node. In other words, it corresponds to the i-th dual photon.

\item Ungauge the neighbouring nodes of the quiver. This results in a `local quiver theory' with a single gauge node. 
Let us call this theory B$_{\rm loc}$.

\item Perform mirror symmetry on this `local quiver theory' B$_{\rm loc}$, calling the resulting theory A$_{\rm loc}$. 
The monopole deformation term will be mapped to an off-diagonal mass term for the matter fields in A$_{\rm loc}$.

\item Integrate out the massive fields in A$_{\rm loc}$, leading to an effective theory ${\tilde {\rm A}_{\rm loc}}$.
\item Compute the mirror of ${\tilde {\rm A}_{\rm loc}}$, which we call ${\tilde {\rm B}_{\rm loc}}$.

\item Couple ${\tilde {\rm B}_{\rm loc}}$ back into the original quiver, by trading the i-th node for it.

\end{enumerate}

Two comments are in order. A generic T-brane corresponds to turning on superpotential terms involving monopole operators at multiple 
nodes. Our strategy is perfectly applicable in this case as well, since we simply need to reiterate the above steps at each node.
This method is particularly effective in the class of theories we are considering because every gauge node is balanced (the 
number of flavors is twice the number of colors). Infact, only in this case the mirror of a monopole operator is a mass term and 
this allows us to simplify the answer by integrating out massive fields.

This should in principle be applicable for the whole ADE series and in the next section we will see that indeed it does work for
the {\bf A} series. For the {\bf D} series, we will focus on Abelian nodes in this paper. This choice can always be made if we assume a 
T-brane along a minimal nilpotent orbit in D$_N$. We leave the study of more general orbits for future work.

\subsection{Local mirror symmetry in the Abelian case}  \label{sec:localmirrorabelian}

Let us discuss how our procedure of `local mirror symmetry' works in the Abelian case. The superpotential for the 
$\mathcal{N}=4$ theory with $N$ $U(1)$ gauge nodes is 
\be W=\sum_{i=1}^N S_iq_i\tilde{q}^i,\ee 
where the $S_i$ multiplets satisfy the constraint $\sum_iS_i=0$. 
It is more convenient to work in terms of unconstrained fields and introduce a dynamical lagrange multiplier. The superpotential 
is then rewritten as follows: 
\be W=\sum_{i=1}^N S_i q_i\tilde{q}^i-\Psi(\sum_{i=1}^N S_i)\:.\ee 
As before, this sets $q_i\tilde{q}^i=\Psi$ for 
every $i$, which is the correct chiral ring relation. 

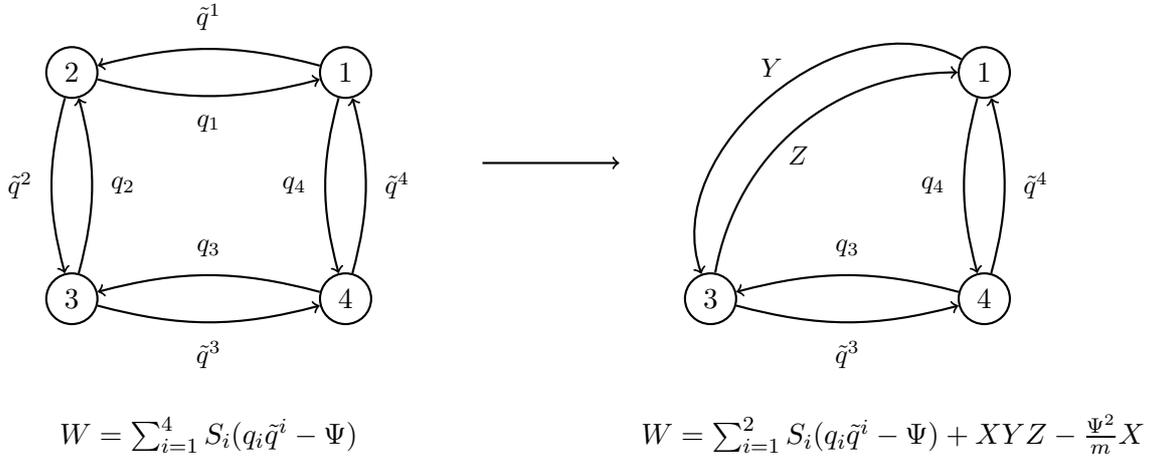
\begin{figure}[ht!] 
\centering
\begin{tikzpicture}[->, every node/.style={circle,draw},thick, scale=0.6]

  \node(M1) at (-17,-3){$3$};

 \node(M2) at (-11,-3){$4$};
    \node(M3) at (-11,2){$1$};
        \node(M4) at (-17,2){$2$};
            \node[draw=none](T1) at (-14,-6) {$W = \sum_{i=1}^4 S_i (q_i \tilde q^i-\Psi)$};

  \node(L1) at (-3,-3){$3$};
  \node(L2) at (3,-3){$4$};
    \node(L3) at (3,2){$1$};
    \node[draw=none](T2) at (1,-6) {$W = \sum_{i=1}^2 S_i (q_i \tilde q^i-\Psi)+XYZ-\frac{\Psi^2}{m} X$};

\draw[=>] (-8,0)--  (-5,0);

 \path[every node/.style={font=\sffamily\small,
  		fill=white,inner sep=1pt}]
(M1) edge [bend right=15] node[below=2mm] {$\tilde q^3$} (M2)
(M2) edge [bend right=15]  node[above=2mm] {$q_3$}(M1)
(M2) edge [bend right=15] node[right=2mm] {$\tilde q^4$} (M3)
(M3) edge [bend right=15] node[left=2mm] {$q_4$}(M2)
(M3) edge [bend right=15] node[above=2mm] {$\tilde q^1$} (M4)
(M4) edge [bend right=15] node[below=2mm] {$q_1$}(M3)
(M4) edge [bend right=15] node[left=2mm] {$\tilde q^2$} (M1)
(M1) edge [bend right=15] node[right=2mm] {$q_2$}(M4)

(L1) edge [bend right=15] node[below=2mm] {$\tilde q^3$} (L2)
(L2) edge [bend right=15]  node[above=2mm] {$q_3$}(L1)
(L2) edge [bend right=15] node[right=2mm] {$\tilde q^4$} (L3)
(L3) edge [bend right=15] node[left=2mm] {$q_4$}(L2)
(L3) edge [bend right=70] node[above=2mm] {$Y$} (L1)
(L1) edge [bend left=40] node[below=2mm] {$Z$}(L3)
;

\end{tikzpicture}
\vskip -2.8cm
\caption{Example of the mirror of SQED with $N=4$ flavors with minimal T-brane.}
 \label{fig:N=4->N=3}
\end{figure}

If we turn on a T-brane with a Jordan block of size two, we should add to the superpotential a term proportional to a monopole 
operator for one gauge node of the quiver. Our proposal for understanding the effect of this deformation is then to focus on this 
node, which is a $U(1)$ theory with two flavors and superpotential 
\be W=S_1 q_1\tilde{q}^1+S_2 q_2\tilde{q}^2-\Psi(S_1+S_2)+mW_{2,+} \:,\ee
where the T-brane is along the node $2$.
For simplicity we do not include the other terms $\Psi(S_3+\dots)$ since they play no role in what follows. We will reintroduce 
them at the end of the computation. The mirror of an $\mathcal{N}=2$ $U(1)$ theory with two flavors and no superpotential is well 
known: it is again an Abelian theory with two flavors plus two neutral chiral multiplets $A_1$, $A_2$ and superpotential 
\be A_1Q\tilde{Q}+A_2P\tilde{P}.\ee 
Under the mirror map the diagonal components of the meson matrix $q_i\tilde{q}^i$ are identified with $A_i$ and $W_+$ becomes 
an off-diagonal mass term. By looking at our gauge node as $\mathcal{N}=2$ SQED plus three neutral chirals $S_1$, $S_2$ and $\Psi$ 
with the above superpotential and exploiting the mirror map dictionary, we can immediately find the mirror theory which is again 
SQED with two flavors and superpotential 
\be W=A_1Q\tilde{Q}+A_2P\tilde{P}+S_1A_1+S_2A_2-\Psi(S_1+S_2)+mP\tilde{Q}.\ee
The fields $S_i$, $A_i$, $P$ and $\tilde{Q}$ are now massive and we can integrate them out. We keep instead $\Psi$ until the end since it is 
coupled to other fields in the quiver. The equations of motion identify $S_i$ with the mesons and we are left with 
\be W=-\frac{\Psi^2}{m}Q\tilde{P}.\ee 
The theory ${\tilde {\rm A}_{\rm loc}}$ in the case at hand is SQED with one flavor and the above superpotential. In order to 
complete our analysis, we now derive the mirror of 
this model and ``reconnect'' the resulting theory to the quiver. Since the mirror of SQED with one flavor is the XYZ model, we get 
the theory 
\be W=XYZ-\frac{\Psi^2}{m}X.\ee 
Interpreting now the fields $Y,Z$ as the bifundamentals of the $U(1)\times U(1)$ gauge symmetry of the neighbouring nodes of the quiver, 
we obtain a circular quiver with $N-1$ $U(1)$ nodes and superpotential 
\be W=XYZ+\sum_{i=1}^{N-2}S_i q_i\tilde{q}^i-\Psi(\sum_{i=1}^{N-2}S_i)-\frac{\Psi^2}{m}X.\ee 
This is exactly the theory we have already found in (\ref{nilpot}), with $X$ playing the role of the extra S field. An example is shown in figure \ref{fig:N=4->N=3}.

Let us now briefly discuss the case of more general T-branes. In the case of a Jordan block of size three, after this procedure 
one of the gauge nodes still has a superpotential term involving monopole operators. 
We should then repeat the above process at that node as well. The mirror is again SQED with two flavors and superpotential 
\be W=A_1Q\tilde{Q}+A_2P\tilde{P}+XA_1+SA_2-\frac{\Psi^2}{m}X-\Psi S+mP\tilde{Q}.\ee 
By integrating out $X$, $S$, $A_i$, $P$ and $\tilde{Q}$ we find 
\be-\frac{\Psi^3}{m^2}\tilde{P}Q.\ee 
Again we find a theory with one flavor whose mirror is a variant of the XYZ model. The final 
quiver we are left with is a circular quiver with $N-2$ nodes and superpotential 
\be W=XYZ+\sum_{i=1}^{N-3}S_iq_i\tilde{q}^i-\Psi(\sum_{i=1}^{N-3}S_i)-\frac{\Psi^3}{m^2}X,\ee 
which is again in perfect agreement with our previous findings. Clearly this procedure can be reiterated for Jordan blocks of 
arbitrary size.

\subsection{Local mirror symmetry in the \texorpdfstring{{\bf D} and {\bf E}}{} series}  \label{sec:localmirrorde}

The above ideas can be immediately applied to the {\bf D} and {\bf E} series as well, at least in the case of minimal nilpotent orbits, 
which requires turning on superpotential terms involving monopole operators at Abelian nodes only. The analysis just involves 
the knowledge of mirror symmetry for $\mathcal{N}=2$ Abelian theories, about which already a lot is known. The general case 
requires non-Abelian mirror symmetry, which will not be discussed in the present paper. 

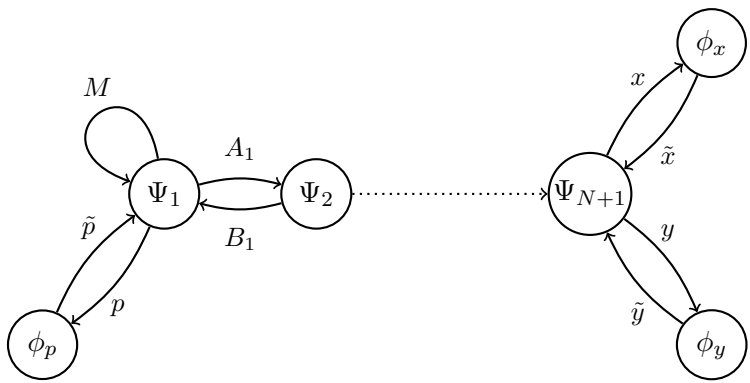
\begin{figure}[ht!]
\begin{center}
\begin{tikzpicture}[->, every node/.style={circle,draw},thick, scale=0.8]
  \node(L2) at (-10,-2.5){$\phi_p$};
  \node(V1) at (-8,0){$\Psi_1$};
  \node(V2) at (-5.5,0){$\Psi_2$};
  \node[inner sep=.7](VN+1) at (-1,0){$\Psi_{N+1}$};
  \node(L3) at (1,-2.5) {$\phi_y$};
    \node(L4) at (1,2.5) {$\phi_x$};

 \path[every node/.style={font=\sffamily\small,
  		fill=white,inner sep=1pt}]
(L2) edge [bend left=15] node[above=2mm] {$\tilde p$} (V1)
(V1) edge [bend left=15] node[below=2mm] {$p$}(L2)
(V1) edge [bend left=15] node[above=2mm] {$A_1$} (V2)
(V2) edge [bend left=15] node[below=2mm] {$B_1$} (V1)
(V2) edge [dotted] (VN+1)
(L3) edge [bend left=15] node[below=2mm] {$\tilde y$} (VN+1)
(VN+1) edge [bend left=15] node[above=2mm] {$y$}(L3)
(L4) edge [bend left=15] node[below=2mm] {$\tilde x$} (VN+1)
(VN+1) edge [bend left=15] node[above=2mm] {$x$}(L4)
(V1) edge[loop, out=100, in=160, looseness=8] node[above=2mm]{$M$} (V1)
;

\end{tikzpicture}
\caption{Effective quiver for deformed D$_{N+4}$ with $ W=-\phi \, \tr M+\tr(\Psi_1 M)-\frac{X}{m} {\det M}+\ldots$}
\label{fig:dn->reduced}
\end{center}
\end{figure}

Let us consider the affine D$_N$ quiver which has four Abelian tails coupled to a $U(2)$ gauge group. In the presence of a 
T-brane related to a minimal nilpotent orbit, we can focus on one of the Abelian tails. If we choose to focus on the node $q$ (see Figure \ref{DNquiver}), the relevant superpotential terms are\footnote{Here the indices of $q$ and $\tilde{q}$ are gauge indices from the full $D_N$ point of view, but are flavor indices if we focus only on the node $q$.} 
\be W=-\phi(q_1\tilde{q}^1+q_2\tilde{q}^2)+\sum_{a=1}^2\tr(\Psi q_a\tilde{q}^a )+mW_{q,+}\:.\ee
In the above formula $\Psi$ is the chiral multiplet in the adjoint of $U(2)$. We now apply the `local mirror symmetry' procedure 
at this node: we have SQED with 2 flavors, so its mirror is again the same type of 
theory. The diagonal components of the meson matrix are mapped to fundamental fields on the mirror side, which we call $s_1$ and 
$s_2$, whereas the off-diagonal components are mapped to monopole operators $w_+$ and $w_-$. The fields $\phi$ and $\Psi$ are 
gauge invariant fields which will be merely spectators in what follows. They have a counterpart in the mirror theory which we will 
again call $\phi$ and $\Psi$. Calling $Q$ and $P$ the flavors on the mirror side, we get the superpotential 
\be\label{dneff} W=-\phi(s_1+s_2)+\tr(\Psi M)+s_1Q\tilde{Q}+s_2P\tilde{P}+mP\tilde{Q},\ee 
where $M$ is a matrix transforming in the adjoint of $U(2)$, whose components are $s_i$ and $w_{\pm}$: 
\be M\equiv\left(\begin{array}{cc}
  s_1 & w_+ \\             
  w_- & s_2 \end{array}\right).\ee 
We now simply integrate out the massive fields $P$ and $\tilde{Q}$, getting an Abelian theory with one flavor. The fields 
$w_{\pm}$ are now interpreted as the monopole operators of the latter. The resulting superpotential is 
\be W=-\phi\tr M+\tr(\Psi M)-\frac{s_1s_2}{m}Q\tilde{P}.\ee
As we have already explained, the mirror of SQED with one flavor is the XYZ model, where X, Y and Z are respectively the mirror 
of the meson and monopole operators. Mirroring again we then find\footnote{The unusual factor of $1/m$ in front of the $XYZ$ term can be 
derived as follows: As was explained in \cite{Aharony:1997bx}, the moduli space of an Abelian theory can be studied by treating the 
monopoles and mesons as elementary fields and supplementing the superpotential with the term $-N_f(w_+w_-\text{det}\mathcal{M})^{1/N_f}$. 
In the case at hand $\mathcal{M}$ is the meson matrix built out of $P$ and $Q$ fields. The superpotential (\ref{dneff}) becomes 
$$s_1\mathcal{M}_{11}+s_2\mathcal{M}_{22}+m\mathcal{M}_{12}-2\sqrt{w_+w_-\text{det}\mathcal{M}}+\dots$$ 
where dots denotes the first two terms in \ref{dneff}. Using just the $\mathcal{M}_{ab}$ F-terms, we can rewrite it as 
$$-\frac{\mathcal{M}_{21}}{m}s_1s_2+\frac{\mathcal{M}_{21}}{m}w_+w_-+\dots$$ 
The first term is precisely what we get integrating out $P$ and $\tilde{Q}$ and the second is the weighted $XYZ$ term. In our 
discussion $X$ is the mirror of $\mathcal{M}_{21}$ whereas $Y$ and $Z$ are to be understood as the mirrors of $w_{\pm}$, which 
are the fields appearing in the matrix $M$.}
\be\label{dienne1} W=-\phi\tr M+\tr(\Psi M)-\frac{s_1s_2}{m}X+\frac{XYZ}{m}.\ee
Since $Y$ and $Z$ are now identified with the off-diagonal components of the field $M$, this can be rewritten as 
\be\label{dienne2} W=-\phi\tr M+\tr(\Psi M)-\frac{X}{m}\text{det}M.\ee Notice that all the above terms are $U(2)$ invariant.
We now glue again our theory to the $U(2)$ gauge node. Since the gauge group has now disappeared, our quiver has lost one Abelian 
tail and has now the shape of a D$_N$ (not affine) Dynkin diagram. The previously trivalent vertex now has two adjoint chiral multiplets 
and two neutral chirals ($\phi$ and $X$) coupled to them. The rest of the quiver and superpotential terms are unaltered. This is illustrated in figure \ref{fig:dn->reduced}.

Armed with this result, we can now make an important observation: As in the A$_{N-1}$ case the Higgs branch is not modified by the 
T-brane. As we have already explained, the Higgs branch of the $\mathcal{N}=4$ theory is the singularity of type D$_N$. 
This can be shown by constructing suitable gauge invariant operators out of the bifundamentals and using the F-term constraints to 
show that they satisfy the desired relation \cite{Lindstrom:1999pz}. The theory we are discussing differs from this more supersymmetric model only in 
one aspect: one of the $U(1)\times U(2)$ bifundamentals is missing. However, all the gauge invariants considered in extracting the 
singularities are constructed using the meson matrix built out of these bifundamentals, and our theory has a perfectly good candidate 
to replace the missing meson: the field $M$ in the adjoint of $U(2)$ introduced above. 

In the $\mathcal{N}=4$ case, the fact that the meson is bilinear in $U(1)\times U(2)$ bifundamentals immediately implies that it 
has rank one (or equivalently the determinant is zero) and the F-term equation associated with the chiral multiplet 
sitting in the $\mathcal{N}=4$ Abelian vectormultiplet tells us that the meson is traceless. These are the only two properties 
needed in extracting the singularities, together with the F-term equations of the various $U(2)$ vectormultiplets which are 
automatically included in our model as well since the relevant superpotential terms are the same. 

A priori, our field $M$ is a generic $2\times 2$ matrix. However, the traceless and zero determinant constraints are implemented 
by the F-term equations of $\phi$ and $X$, as it is clear from (\ref{dienne2}). This guarantees that the vev of $M$ can 
be written in the form $M=q\tilde{q}$, with $q$ and $\tilde{q}$ two dimensional vectors satisfying the same constraints as the 
bifundamentals in 
the $\mathcal{N}=4$ theory. Hence, we can straightforwardly repeat the argument valid for the more supersymmetric case and 
conclude that our D$_N$-shaped quiver reproduces the D$_N$ singularity. Clearly the procedure can be repeated for other Abelian 
tails as well, with exactly the same conclusion. For example, we could turn on monopole superpotential terms at all the Abelian 
nodes and get a linear quiver (the gauge nodes at the two ends have two flavors and three adjoints) which again reproduces the D$_N$ singularity. 

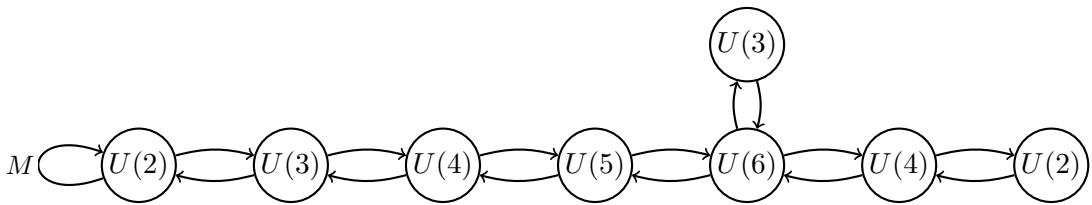
\begin{figure}[ht!]
\begin{center}
\begin{tikzpicture}[->, every node/.style={circle,draw},thick, scale=0.8]
  \node[inner sep=1](V1) at (-8,0){$U(2)$};
  \node[inner sep=1](V2) at (-5.5,0){$U(3)$};
  \node[inner sep=1](V3) at (-3,0){$U(4)$};
  \node[inner sep=1](V4) at (-.5,0){$U(5)$};
  \node[inner sep=1](V5) at (2,0){$U(6)$};
  \node[inner sep=1](V6) at (4.5,0){$U(4)$};
    \node[inner sep=1](V7) at (7,0){$U(2)$};
  \node[inner sep=1](V8) at (2,2){$U(3)$};

 \path[every node/.style={font=\sffamily\small,
  		fill=white,inner sep=1pt}]
(V1) edge [bend left=15]   (V2)
(V2) edge [bend left=15]   (V1)
(V2) edge [bend left=15]   (V3)
(V3) edge [bend left=15]   (V2)
(V3) edge [bend left=15]   (V4)
(V4) edge [bend left=15]   (V3)
(V4) edge [bend left=15]   (V5)
(V5) edge [bend left=15]   (V4)
(V5) edge [bend left=15]   (V6)
(V6) edge [bend left=15]   (V5)
(V6) edge [bend left=15]   (V7)
(V7) edge [bend left=15]   (V6)
(V5) edge [bend left=15]   (V8)
(V8) edge [bend left=15]   (V5)
(V1) edge [loop, out=200, in=160, looseness=9]  node[left]{$M$}  (V1)

;

\end{tikzpicture}
\caption{Effective `mutilated' quiver for E$_8$  with $ W=-\phi \, \tr M+\tr(\Psi_{U(2)} M)-\frac{X}{m} {\det M}+\ldots$}\label{Fig:E8}
\end{center}
\end{figure}

The same analysis can be repeated straightforwardly for the {\bf E} series as well: the $U(1)$ node is replaced by an adjoint and 
two neutral chirals for the neighbouring $U(2)$ node (the superpotential is again as in \eqref{dienne2}) and the E$_N$ singularity of the $\mathcal{N}=4$ theory will be preserved. 
In this way we can e.g. find a candidate for the mirror of the dimensional reduction of Minahan-Nemeshansky theories, deformed 
by a minimal nilpotent orbit mass term. This is displayed for the $E_8$ case in Figure \ref{Fig:E8}. Here, the minimal T-brane removes the $U(1)$ node of the E$_8$ quiver. This node is associated to the highest root. By a different choice of basis, this root can be mapped to any (simple) root, proving that one can treat each element of the minimal nilpotent orbit.

\section{Conclusions and outlook}\label{sec:conclusions}
In this paper, we find the mirror theory for D2-branes probing a stack of T-branes. On the A-side, this corresponds to an off-diagonal mass deformation. On the B-side, it corresponds to deforming the superpotential via monopole operators. This provides us with a definition of a T-brane directly in terms of a membrane probing a singularity, even for the {\bf E} series. The uplift from a D2 to an M2-brane probing the singularity means flowing to the IR fixed point.

By using a technique we dubbed `local mirror symmetry', meaning performing mirror symmetry on a single node of the quiver at a time, we were able to study T-branes along minimal nilpotent orbits, for any ADE singularity. The result is that the effective theory is described by a reduced quiver which has the same Higgs branch as the original quiver. 

The problem of studying generic nilpotent orbits is more difficult, as it requires understanding non-Abelian $\cN=2$ mirror symmetry, which is not only technically difficult, but also prone to instanton corrections. Nevertheless, it would be interesting to pursue this further.

A related puzzle is the following: Modulo a Weyl transformation the effect of a minimal T-brane is described by a monopole superpotential 
term at a single gauge node; any node is fine and the $\mathcal{N}=2$ theories obtained by different choices are equivalent. 
In the case of the {\bf A} series this is obvious, since all the gauge nodes are equivalent. In the case of {\bf D} and {\bf E} 
theories on the other hand, this leads to the prediction that by turning on a specific monopole deformation at a non-Abelian node (the monopole should be the one paired by supersymmetry with the current corresponding to the root associated with the node) we get 
a theory which is dual to those described in the previous section. From the field theory perspective this is a rather surprising 
statement and it would certainly be interesting to elucidate this point. We hope to come back to this issue in the future.

It would also be interesting to derive formulae for the modified Coulomb branches of our deformed B-theories. Perhaps there might be a way to amend the Hilbert series constructions of \cite{Benini:2009qs,Cremonesi:2014kwa,Cremonesi:2013lqa,Cremonesi:2015dja} that proved so successful in constructing moduli spaces. Hilbert series may also help to improve our knowledge of the $\cN=2$ version of non-Abelian mirror symmetry, perhaps providing a way of inferring the mirror dual of operators like $\tilde{Q}\Phi Q$, which, in contrast to the $\cN=4$ case, are non-trivial in the chiral ring.

Another open question is the following: In $\cN=4$ theories, monopole operators come in multiplets that contain (spin one) conserved currents. This yields a powerful method to derive the quantum enhanced flavor symmetry of quiver gauge theories by simply looking for monopole operators of R-charge one via a zero-mode counting technique. For $\cN=2$ theories the link between monopole operators and conserved currents is in principle lost. Nevertheless, in the class of theories we studied (in the {\bf A} and {\bf D} series), we can make predictions about the global symmetries through mirror symmetry. It would be a significant step forward if techniques were developed to find these directly in $\cN=2$ quiver gauge theories without resort to mirror symmetry.

\section*{Acknowledgements}
We would like to extend our gratitude to F. Benini, C. Closset, S. Cremonesi, M~ Fazzi, D.~Forcella, N.~Mekareeya and A. Zaffaroni for useful discussions.

A.C. is a Research Associate of the Fonds de la Recherche Scientifique F.N.R.S. (Belgium). The work of A.C. is partially supported by IISN - Belgium (convention 4.4503.15). 
S.G. is partially supported by the ERC Advanced Grant ``SyDuGraM", by FNRS-Belgium (convention FRFC PDR T.1025.14 and 
convention IISN 4.4503.15) and by the ``Communaut\'e Fran\c{c}aise de Belgique" through the ARC program.
The work of R.S. is supported by the ERC Starting Independent Researcher Grant 259133- ObservableString.
The work of A.C., R.S. and R.V. was performed in part at the Aspen Center for Physics, which is supported by National Science Foundation grant PHY-1066293.

\appendix
\section{Mirror map for \texorpdfstring{D$_4$}{} theories}\label{AppendixA}

The mirror map acts on the meson matrix defining HB$_A$ by sending it to a matrix whose elements are coordinates on CB$_B$, i.e. monopole operators and combinations of the scalar fields $\Psi$, $\phi_q$, $\phi_x$, $\phi_y$, $\phi_p$, where we have only one $U(2)$ node. We keep the form \eqref{MorietifoldForm} for the meson matrix $M$. We write $\Psi=\phi_t \mathbb{1} + \psi$, where $\phi_t$ is the component along the diagonal $U(1)$ of $U(2)$ and $\psi=\psi_t\sigma_3+\psi_1 \sigma_1+\psi_2\sigma_2$ is in the adjoint of $SU(2)$ ($\sigma_i$ are the Pauli matrices).

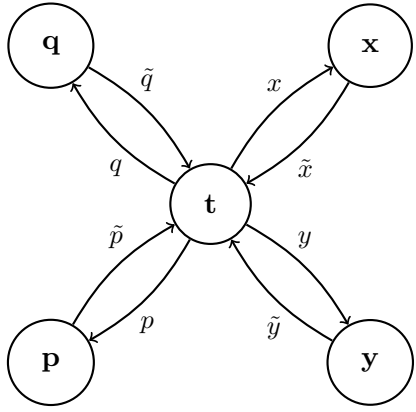
\begin{figure}
\begin{center}
\begin{tikzpicture}[->, every node/.style={circle,draw},thick, scale=0.7]
  \node[inner sep=1](L1) at (-6,3){$\,\,\,\,\,\,\bf q\,\,\,\,\,\,$};
  \node[inner sep=1](L2) at (-6,-3){$\,\,\,\,\,\,\bf p\,\,\,\,\,\,$};
  \node[inner sep=1](V1) at (-3,0){$\,\,\,\,\,\,\bf t\,\,\,\,\,\,$};
  \node[inner sep=1](L3) at (0,-3) {$\,\,\,\,\,\,\bf y\,\,\,\,\,\,$};
    \node[inner sep=1](L4) at (0,3) {$\,\,\,\,\,\,\bf x\,\,\,\,\,\,$};

 \path[every node/.style={font=\sffamily\small,
  		fill=white,inner sep=1pt}]
(L1) edge [bend left=15] node[above=2mm] {$\tilde q$} (V1)
(V1) edge [bend left=15] node[below=2mm] {$q$}(L1)
(L2) edge [bend left=15] node[above=2mm] {$\tilde p$} (V1)
(V1) edge [bend left=15] node[below=2mm] {$p$}(L2)
(L3) edge [bend left=15] node[below=2mm] {$\tilde y$} (V1)
(V1) edge [bend left=15] node[above=2mm] {$y$}(L3)
(L4) edge [bend left=15] node[below=2mm] {$\tilde x$} (V1)
(V1) edge [bend left=15] node[above=2mm] {$x$}(L4)

;

\end{tikzpicture}
\end{center}
\caption{D$_4$ quiver. The quiver labels in boldface represent monopole charges.} \label{fig:d4quiver}
\end{figure}

The diagonal elements of the meson matrix are mapped to some combinations of the scalars that live in the $U(1)$ vector multiplet. This is done by mapping diagonal mass terms to FI-terms. The off-diagonal elements are  mapped to monopoles operators with R-charge 1 on the B-side (see \cite{Bashkirov:2010hj}). To see which one maps to which, one needs to compute the charges of  the off-diagonal elements of the meson-matrix with respect to the mirror of  topological $U(1)$'s relative to the nodes of the D$_N$ quiver on the B-side. We call the  charges of these topological symmetries $\bf (t,q,x,y)$, following the diagram in figure \ref{fig:d4quiver}. 
On the A-side one easily compute the charges of the off-diagonal elements of $M$. So we can write down $M$ by substituting off diagonal mesons with the corresponding monopole operators $v_{tqxy}$. Here a gauge fixing has been done, such that the charge with respect to the $p$-node is zero ($p=0$) \cite{Bashkirov:2010hj}. 

The mirror map can then be written as
\begin{equation}
M = \left( \begin{array}{ccc}
 Q \cdot \tilde{Q}   &&   Q \cdot \epsilon \cdot  Q^T   \\  \\
  \tilde{Q}^T \cdot\epsilon^{-1}\cdot \tilde{Q}   &&   - \tilde{Q}^T\cdot Q^T  \\
\end{array}\right)\nonumber
\end{equation}
\begin{equation}
\updownarrow 
\end{equation}
\begin{equation}
\left(\begin{array}{cccccccc}
\alpha_1  & v_{0100}  & v_{1100} & v_{1110} & 0 & v_{2111}  & v_{1111} & v_{1101} \\
v_{0\mbox{-}100} & \alpha_2 & v_{1000} & v_{1010} & -v_{2111} & 0 & v_{1011} & v_{1001}  \\
v_{\mbox{-}1\mbox{-}100} & v_{\mbox{-}1000} & \alpha_3 & v_{0010} &  -v_{1111} &  -v_{1011} & 0 & v_{0001}  \\
v_{\mbox{-}1\mbox{-}1\mbox{-}10} & v_{\mbox{-}10\mbox{-}10}  & v_{00\mbox{-}10} & \alpha_4 &  -v_{1101} &  -v_{1001} & -v_{0001} & 0  \\
 0 & - v_{\mbox{-}2\mbox{-}1\mbox{-}1\mbox{-}1}  & -v_{\mbox{-}1\mbox{-}1\mbox{-}1\mbox{-}1} & -v_{\mbox{-}1\mbox{-}10\mbox{-}1} & -\alpha_1  & -v_{0\mbox{-}100}  & -v_{\mbox{-}1\mbox{-}100} & -v_{\mbox{-}1\mbox{-}1\mbox{-}10}  \\
 v_{\mbox{-}2\mbox{-}1\mbox{-}1\mbox{-}1} & 0 & -v_{\mbox{-}10\mbox{-}1\mbox{-}1} & -v_{\mbox{-}100\mbox{-}1}  & -v_{0100} & -\alpha_2 & -v_{\mbox{-}1000} & -v_{\mbox{-}10\mbox{-}10}  \\
 v_{\mbox{-}1\mbox{-}1\mbox{-}1\mbox{-}1} &  v_{\mbox{-}10\mbox{-}1\mbox{-}1} & 0 & -v_{000\mbox{-}1}  & -v_{1100} &- v_{1000} & -\alpha_3 & -v_{00\mbox{-}10} \\
 v_{\mbox{-}1\mbox{-}10\mbox{-}1} &  v_{\mbox{-}100\mbox{-}1} & v_{000\mbox{-}1} & 0 & -v_{1110} & -v_{1010}  & -v_{0010} & -\alpha_4  \\
\end{array}\right)\:, \nonumber
\end{equation}
where we have implemented the orientifold conditions on $Q,\tilde{Q}$ to bring $M$ in the form \eqref{MorietifoldForm} ($Q$ is meant to be a $N\times 2$ matrix, while $\tilde{Q}$ is a $2\times N$ matrix; moreover $\epsilon_{ab}\equiv i\gamma_{ab}$ while $\epsilon^{ab}\equiv-i\gamma^{ab}$).
The diagonal elements $\alpha_1,\alpha_2,\alpha_3,\alpha_4$ 
on the B side are\footnote{They can be found by mapping appropriately diagonal mass terms in theory A to FI-terms in theory B.}
\begin{equation}\label{alpha1234}
 (\alpha_1,\alpha_2,\alpha_3,\alpha_4) = (\phi_p - \phi_q, 2\phi_t-\phi_p-\phi_q, 2\phi_t-\phi_x-\phi_y, \phi_y-\phi_x)\:,
\end{equation}
where $\vec{\phi}=(\phi_t+\psi_t,\phi_t-\psi_t,\phi_q,\phi_x,\phi_y,\phi_p)$ are the scalars in the vector multiplets on the corresponding nodes (the first ones correspond to the $U(1)\times U(1)$ inside $U(2)$ of the central node).

We want to show that applying this map and the quantum relations between monopole operators, one is able to find on the B-side the vanishing of the $4\times 4$ minors. This implies that $M$ has rank 2 and would verify the mirror map itself.
The rules we need are \cite{Bullimore:2015lsa}:
\begin{equation}\label{GaiottovAvB}
 v_A v_B = v_{A+B} \frac{P^{\rm hyp}(\vec{\phi})}{P^{\rm W}(\vec{\phi})}
\end{equation}
where\footnote{We are using here a different convention with respect to \cite{Bullimore:2015lsa}: our monopole operators $v_A$ are multiplyed by $(-i)^{|t_A|}$ with respect to the ones appearing in \cite{Bullimore:2015lsa}.}
\begin{eqnarray}
P^{\rm hyp}(\vec{\phi}) &=& i^{|t_{A+B}|}\prod_{i={\rm hyp}} \langle \mu_i,\vec{\phi} \rangle^{\langle\mu_i,A\rangle_+ + \langle\mu_i,B\rangle_+ - \langle\mu_i,A+B\rangle_+} \\
P^{\rm W}(\vec{\phi}) &=& (-i)^{|t_A|+|t_B|} \prod_{j={\rm roots}} \langle \alpha_j,\vec{\phi} \rangle^{\langle\alpha_j,A\rangle_+ + \langle\alpha_j,B\rangle_+ - \langle\alpha_j,A+B\rangle_+}
\end{eqnarray}
$A,B$ are the charge vectors that select the monopole operator in the Abelianized theory, i.e. in the $U(1)^6$ theory that lives along the Coulomb branch. In our case $A,B=(t_1,t_2,q,x,y,p)$. $\mu_i$ are the charge vector of the hypermultiples; in our case
\begin{eqnarray}
 \mu_{y-t_1} = (1,0,0,0,-1,0)   &&  \mu_{y-t_2} = (0,1,0,0,-1,0) \\
 \mu_{x-t_1} = (1,0,0,-1,0,0)   &&  \mu_{x-t_2} = (0,1,0,-1,0,0) \\
  \mu_{q-t_1} = (1,0,-1,0,0,0)   &&  \mu_{q-t_2} = (0,1,-1,0,0,0) \\
 \mu_{p-t_1} = (1,0,0,0,0,-1)   &&  \mu_{p-t_2} = (0,1,0,0,0,-1) 
\end{eqnarray}
$\alpha_j$ are the charges of the roots
\begin{eqnarray}
 \alpha_{+} = (1,-1,0,0,0,0)   &&  \alpha_{-} = (-1,1,0,0,0,0) \:.
\end{eqnarray}
Finally $\langle V_1 , V_2 \rangle_+$ is defined to be zero if  $\langle V_1 , V_2 \rangle$ is negative and equal to $\langle V_1 , V_2 \rangle$ (the euclidean scalar product of the two vectors)  if it is positive.
We start computing the masses $\langle \mu, \vec{\phi} \rangle$ of the hypermultiplets:
\begin{eqnarray}
 \langle \mu_{\ell-t_1},\vec{\phi} \rangle  = \phi_t + \psi_t -\phi_\ell    &&   \langle \mu_{\ell-t_2},\vec{\phi} \rangle  = \phi_t - \psi_t -\phi_\ell  
\end{eqnarray}
with $\ell=p,q,x,y$. The massess of the two roots are 
\begin{eqnarray}
 \langle \alpha_{+},\vec{\phi} \rangle  = 2 \psi_t     &&   \langle \alpha_{-},\vec{\phi} \rangle  =  - 2 \psi_t \:.
\end{eqnarray}

We now apply the formula \eqref{GaiottovAvB} to the case of interest. The charges of the monopoles operators in the Abelianized theory are $(t_1,t_2,q,x,y,p)$ with $p=0$ and $t_1+t_2=t$.
Note that the monopoles operators with $t=0$ are easily defined on the B-side, as they are charged only under the $U(1)$ nodes. 
We start from a minor that includes only these types of operators:
\begin{equation}\label{simpleMinor}
\left| \begin{array}{cccc}
 \alpha_3 & v_{0010} & 0 & v_{0001} \\
  v_{00\mbox{-}10} & \alpha_4 & - v_{0001} & 0 \\
 0 & -v_{000\mbox{-}1} & -\alpha_3 & -v_{00\mbox{-}10} \\
   v_{000\mbox{-}1} & 0 & -v_{0010} & -\alpha_4 \\
\end{array}    \right| =  \left( \alpha_3\alpha_4 + v_{0001}v_{000\mbox{-}1} - v_{0010}v_{00\mbox{-}10} \right)^2
\end{equation}
By using \eqref{GaiottovAvB}, we can compute $ v_{0001}v_{000\mbox{-}1}$ and $v_{0010}v_{00\mbox{-}10}$:
\begin{eqnarray}
v_{0001}v_{000\mbox{-}1} &=& (\phi_t+\psi_t-\phi_y)(\phi_t-\psi_t-\phi_y) = (\phi_t-\phi_y)^2- \psi_t^2 \\
v_{0010}v_{00\mbox{-}10} &=& (\phi_t+\psi_t-\phi_x)(\phi_t-\psi_t-\phi_x) = (\phi_t-\phi_x)^2- \psi_t^2 
\end{eqnarray}
Hence 
\begin{equation}
v_{0001}v_{000\mbox{-}1} - v_{0010}v_{00\mbox{-}10} = (\phi_t-\phi_y)^2 - (\phi_t-\phi_x)^2 = - (\phi_y-\phi_x)(2\phi_t-\phi_x-\phi_y)
\end{equation}
that is consistent with the vanishing of \eqref{simpleMinor}, since $\alpha_3\alpha_4 = (2\phi_t-\phi_c-\phi_d) (\phi_d-\phi_c)$.

\

Now, let us consider the minor
\begin{equation}\label{notsosimpleMinor}
\left| \begin{array}{cccc}
 \alpha_1 & v_{1100} & 0 & v_{1111} \\
  v_{\mbox{-}1\mbox{-}100} & \alpha_3 & - v_{1111} & 0 \\
 0 & -v_{\mbox{-}1\mbox{-}1\mbox{-}1\mbox{-}1} & -\alpha_1 & -v_{\mbox{-}1\mbox{-}100} \\
   v_{\mbox{-}1\mbox{-}1\mbox{-}1\mbox{-}1} & 0 & -v_{1100} & -\alpha_3 \\
\end{array}    \right| =  \left( \alpha_1\alpha_3 + v_{1111}v_{\mbox{-}1\mbox{-}1\mbox{-}1\mbox{-}1} - v_{1100}v_{\mbox{-}1\mbox{-}100} \right)^2
\end{equation}
We see that the monopole operators involved have $t=1$. We associate to these monopoles the gauge invariant sum of those with charges  $(t_1,t_2)=(1,0)$ and $(t_1,t_2)=(0,1)$. For example:
\begin{equation}
 v_{1100}\equiv v_{101000} + v_{011000}
\end{equation}
where on the right hand side we have written all the charges ($A,B$) with respect to the Abelianized theory, i.e. $(t_1,t_2,q,x,y,p)$.
Let us start by computing $v_{1100}v_{\mbox{-}1\mbox{-}100}$:
\begin{eqnarray}
v_{1100}v_{\mbox{-}1\mbox{-}100} &=& (v_{101000} + v_{011000}) (v_{\mbox{-}10\mbox{-}1000} + v_{0\mbox{-}1\mbox{-}1000} ) \\
   &=& (v_{101000}v_{\mbox{-}10\mbox{-}1000} + v_{011000}v_{0\mbox{-}1\mbox{-}000}) + ( v_{011000}v_{\mbox{-}10\mbox{-}1000} +v_{101000} v_{0\mbox{-}1\mbox{-}1000} ) \nonumber\\
   &\equiv& \mathcal{O}_{000000} + \mathcal{O}_{1\mbox{-}10000}
\end{eqnarray}
Let us do it step by step by using \eqref{GaiottovAvB}
\begin{eqnarray}
v_{101000}v_{\mbox{-}10\mbox{-}1000} &=& \frac{(\phi_t+\psi_t-\phi_y)(\phi_t+\psi_t-\phi_x)(\phi_t-\psi_t-\phi_q)(\phi_t+\psi_t-\phi_p)}{4\psi_t^2}\\
v_{011000}v_{0\mbox{-}1\mbox{-}1000} &=& \frac{(\phi_t-\psi_t-\phi_y)(\phi_t-\psi_t-\phi_x)(\phi_t+\psi_t-\phi_q)(\phi_t-\psi_t-\phi_p)}{4\psi_t^2}\\
v_{011000}v_{\mbox{-}10\mbox{-}1000} &=& v_{\mbox{-}110000} \\
v_{101000}v_{0\mbox{-}1\mbox{-}000} &=& v_{1\mbox{-}10000} \\
\end{eqnarray}

Analogously
\begin{eqnarray}
v_{1111}v_{\mbox{-}1\mbox{-}1\mbox{-}1\mbox{-}1} &=& (v_{101110} + v_{011110}) (v_{\mbox{-}10\mbox{-}1\mbox{-}1\mbox{-}10} + v_{0\mbox{-}1\mbox{-}1\mbox{-}1\mbox{-}10} ) \\
   &=& (v_{101110}v_{\mbox{-}10\mbox{-}1\mbox{-}1\mbox{-}10} + v_{011110}v_{0\mbox{-}1\mbox{-}1\mbox{-}1\mbox{-}10}) + ( v_{011110}v_{\mbox{-}10\mbox{-}1\mbox{-}1\mbox{-}10} +v_{101110} v_{0\mbox{-}1\mbox{-}1\mbox{-}1\mbox{-}10} ) \nonumber\\
   &\equiv& \mathcal{O}_{000000}' + \mathcal{O}_{1\mbox{-}10000}'
\end{eqnarray}
Again, step by step:
\begin{eqnarray}
v_{101110}v_{\mbox{-}10\mbox{-}1\mbox{-}1\mbox{-}10} &=& \frac{(\phi_t-\psi_t-\phi_y)(\phi_t-\psi_t-\phi_x)(\phi_t-\psi_t-\phi_q)(\phi_t+\psi_t-\phi_p)}{4\psi_t^2}\\
v_{011110}v_{0\mbox{-}1\mbox{-}1\mbox{-}1\mbox{-}10} &=& \frac{(\phi_t+\psi_t-\phi_y)(\phi_t+\psi_t-\phi_x)(\phi_t+\psi_t-\phi_q)(\phi_t-\psi_t-\phi_p)}{4\psi_t^2}\\
v_{011110}v_{\mbox{-}10\mbox{-}1\mbox{-}1\mbox{-}10} &=& v_{\mbox{-}110000} \\
v_{101110}v_{0\mbox{-}1\mbox{-}1\mbox{-}1\mbox{-}10} &=& v_{1\mbox{-}10000} \\
\end{eqnarray}
We now need to put everything together. First of all, we notice that $\mathcal{O}_{1\mbox{-}10000}=\mathcal{O}_{1\mbox{-}10000}'$. This means that these pieces cancel in the difference \eqref{notsosimpleMinor}. We now concentrate on $\mathcal{O}_{000000}-\mathcal{O}_{000000}'$:
\begin{eqnarray}
\mathcal{O}_{000000}-\mathcal{O}_{000000}' &=& (v_{101000}v_{\mbox{-}10\mbox{-}1000} - v_{101110}v_{\mbox{-}10\mbox{-}1\mbox{-}1\mbox{-}10} ) + [\psi_t\mapsto - \psi_t] \nonumber \\  \nonumber \\ 
&=& \frac{(\phi_t-\psi_t-\phi_q)(\phi_t+\psi_t-\phi_p)(2\phi_t -\phi_x-\phi_y)}{2\psi_t} + [\psi_t\mapsto - \psi_t] \nonumber \\ 
&=&  (\phi_p-\phi_x)(2\phi_t -\phi_x-\phi_y) 
\end{eqnarray}
This is consistent with the vanishing of \eqref{notsosimpleMinor} since $\alpha_1\alpha_3 =  (\phi_p-\phi_q)(2\phi_t -\phi_x-\phi_y)$.

\

\noindent
Finally, let us consider the minor
\begin{equation}\label{notsosimpleMinoralpha3}
\left| \begin{array}{cccc}
 v_{\mbox{-}1000} & \alpha_3   & 0 & v_{0001} \\
  v_{\mbox{-}10\mbox{-}10} & v_{00\mbox{-}10} &  - v_{0001} & 0 \\
 0 & -v_{\mbox{-}10\mbox{-}1\mbox{-}1} & -v_{\mbox{-}1000}  & -v_{\mbox{-}10\mbox{-}10} \\
   v_{\mbox{-}10\mbox{-}1\mbox{-}1} & 0 & -\alpha_3 & -v_{00\mbox{-}10} \\
\end{array}    \right| =  \left(  v_{\mbox{-}1000} v_{00\mbox{-}10} + v_{0001}v_{\mbox{-}10\mbox{-}1\mbox{-}1} - \alpha_3 v_{\mbox{-}10\mbox{-}10} \right)^2
\end{equation}
Let us start by computing $v_{\mbox{-}1000}v_{00\mbox{-}10}$:
\begin{eqnarray}
v_{\mbox{-}1000}v_{00\mbox{-}10} &=& (v_{\mbox{-}100000} + v_{0\mbox{-}10000}) v_{000\mbox{-}100} 
\end{eqnarray}
By using the relations \eqref{GaiottovAvB}, we obtain
\begin{eqnarray}
v_{\mbox{-}100000}  v_{000\mbox{-}100}  &=& v_{\mbox{-}100\mbox{-}100} (\phi_t+\psi_t-\phi_x)     \\
 v_{0\mbox{-}10000} v_{000\mbox{-}100} &=& v_{0\mbox{-}10\mbox{-}100}   (\phi_t-\psi_t-\phi_x)\:.      
\end{eqnarray}
Analogously
\begin{eqnarray}
v_{00001}v_{\mbox{-}10\mbox{-}1\mbox{-}1} &=&v_{000010} (v_{\mbox{-}100\mbox{-}1\mbox{-}10} + v_{0\mbox{-}10\mbox{-}1\mbox{-}10})  
\end{eqnarray}
and 
\begin{eqnarray}
v_{000010} v_{\mbox{-}100\mbox{-}1\mbox{-}10}   &=& v_{\mbox{-}100\mbox{-}100} (\phi_t-\psi_t-\phi_y)     \\
 v_{000010} v_{0\mbox{-}10\mbox{-}1\mbox{-}10}  &=& v_{0\mbox{-}10\mbox{-}100}   (\phi_t+\psi_t-\phi_y)      
\end{eqnarray}
Putting everything together:
\begin{eqnarray}
v_{\mbox{-}1000} v_{00\mbox{-}10} + v_{0001}v_{\mbox{-}10\mbox{-}1\mbox{-}1} &=& (2\phi_t-\phi_x-\phi_y)( v_{\mbox{-}100\mbox{-}100} + v_{0\mbox{-}10\mbox{-}100} ) \\
 &=& (2\phi_t-\phi_x-\phi_y) v_{\mbox{-}10\mbox{-}10}  \nonumber
\end{eqnarray}
Since $\alpha_3 =  (2\phi_t-\phi_c-\phi_d)$, this is consistent with the vanishing of the minor.

If one continues along these lines, one can check that all the antisymmetric minors $4\times 4$ vanish if we impose the quantum relations \eqref{GaiottovAvB}, implying that the dual of the meson matrix has rank 2. 

This whole precedure can be easily generalized to D$_N$ for generic $N$. The diagonal elements of $M$ will now include the differences 
$\phi_{t_i}-\phi_{t_{i-1}}$ of the $U(2)$ adjacent nodes.

\providecommand{\href}[2]{#2}\begingroup\raggedright\endgroup

\end{document}